\documentclass[11pt,english]{article}
\usepackage[T1]{fontenc}
\usepackage[a4paper]{geometry}
\usepackage[plainpages=false, colorlinks=true, anchorcolor=blue, linkcolor=blue, citecolor=blue, bookmarks=false]{hyperref}
\usepackage{float} 
\usepackage[utf8]{inputenc}
\usepackage{amsmath}
\usepackage{amssymb}
\usepackage{babel}
\usepackage{color}
\usepackage{graphicx} 

\pdfoutput=1

\begin{document}

\title{Testing a varying-$\Lambda$ model for dark energy within Co-varying Physical Couplings framework}

\author{R. R. Cuzinatto$^{1,2}$\footnote{rcuzinat@uottawa.ca; rodrigo.cuzinatto@unifal-mg.edu.br}, R. P. Gupta$^1$\footnote{rgupta4@uottawa.ca}, R. F. L. Holanda$^3$\footnote{holandarfl@fisica.ufrn.br}, J. F. Jesus$^4$\footnote{jf.jesus@unesp.br}, and S. H. Pereira$^5$\footnote{s.pereira@unesp.br}
\\
\\
$^1$Department of Physics, \\ 
University of Ottawa, \\ Ottawa, ON, K1N 6N5, Canada
\\ \\
$^2$Instituto de Ci\^encia e Tecnologia, \\ 
Universidade Federal de Alfenas, \\ Po\c cos de Caldas, MG, 37715-400, Brazil
\\ \\
$^3$Departamento de F\'isica Te\'orica e Experimental, \\Universidade Federal do Rio Grande do Norte, \\Natal, RN, 59300-000, Brazil
\\ \\
$^4$Instituto de Ci\^encias e Engenharia, \\
Universidade Estadual Paulista (UNESP), \\
R. Geraldo Alckmin 519, \\
Itapeva, SP, 18409-010, Brazil
\\ \\
$^5$Departamento de F\'isica, \\
Faculdade de Engenharia de Guaratinguet\'a,\\
Universidade Estadual Paulista (UNESP),  \\ Guaratinguet\'a, SP,  12516-410, Brazil
}

\date{}

\maketitle

\begin{abstract}
The Co-varying Physical Couplings (CPC) framework is a modified gravity set up assuming Einstein Field Equations wherein the quantities $\{G,c,\Lambda\}$ are promoted to space-time functions. Bianchi identity and the requirement of stress-energy tensor conservation entangle the possible variations of the couplings $\{G,c,\Lambda\}$, which are forced to co-vary as dictated by the General Constraint (GC). In this paper we explore a cosmological model wherein $G$, $c$ and $\Lambda$ are functions of the redshift respecting the GC of the CPC framework. We assume a linear parametrization of $\Lambda$ in terms of the scale factor $a$. We use the ansatz $\dot{G}/G = \sigma \left( \dot{c}/c \right)$  with $\sigma =$ constant to deduce the functional forms of $c=c(z)$ and $G=G(z)$. We show that this varying-$\{G,c,\Lambda\}$ model fits SNe Ia data and $H(z)$ data with $\sigma = 3$. The model parameters can be constrained to describe dark energy at the background level.
 
\end{abstract}

\bigskip
\textbf{Key words:} gravitation -- cosmological parameters -- cosmology: theory -- dark energy -- distance scale.
%

\newpage

\tableofcontents{}


\section{Introduction}


There has been a resurgence of the interest in the subject of varying fundamental constants in gravitation, astrophysics and cosmology as evidenced by recent publications (e.g. \cite{aich2022phenomenological,anagnostopoulos2022swiss,ballardini2021cosmological,bonanno2021effective,chakrabarti2022generalized,cuzinatto2022dynamical,cuzinatto2022observational,da2022fundamental,eaves2021constraints,franchino2021cosmological,Gupta:2020wgz,gupta2022faint,gupta2022varying,gupta2022effect,gupta2022constraining,hanimeli2022can,joseph2022cosmology,lee2021determination,lee2021cosmic,lee2021minimally,leon2022inflation,maeda2022cuscuta,martins2021varying,martins2022varying,mendoncca2021search,pitjeva2021estimates,sakr2022can,sonia2022dynamical}) in the recent literature. This is a rather controversial topic  \cite{ellis2005c,uzan2003fundamental,uzan2011varying}. However, the idea of varying couplings have roots in scalar-tensor theories such as Brans-Dicke theory wherein the gravitational coupling is understood as a scalar field \cite{brans1961mach,faraoni2004cosmology}. The fact that Brans-Dicke theory is equivalent to other modified gravity theories \cite{cuzinatto2016scalar,sotiriou2010f}, that it is recovered from particular limits of string-theory \cite{callan1985strings,fradkin1985quantum} and Kaluza-Klein models \cite{bailin1987kaluza,overduin1997kaluza}, and that it might be related to modifications of the underlying geometry of the space-time manifold \cite{Cuzinatto:2021ttc,faraoni2004cosmology} justifies why this possibility would be interesting. On top of that, the standard model of cosmology faces some challenges \cite{abdalla2022cosmology,bull2016beyond}, themselves pointing to possible limitations of general relativity (GR) as the final theory of gravitational interaction. Examples of the referred difficulties include the lack of a consistent fully quantized version of GR (e.g. \cite{carlip2001quantum}), the presence of singularities (e.g. \cite{bojowald2008loop}), the need for an early inflationary phase (e.g. \cite{baumann2009tasi}) and for a late-time dynamics dominated by dark energy (e.g. \cite{bamba2012dark}). Further open problems of relativistic cosmology are the Hubble tension \cite{di2021realm,heisenberg2022simultaneously,lee2022local}, the $S_8$ tension \cite{abdalla2022cosmology}, and discrepancies related to the distribution of cold dark matter \cite{klypin1999missing}.  This context opens up the possibility of analyzing the consequences of admitting varying physical constants both from theoretical side (e.g. \cite{bonanno2021effective,costa2019covariant,cuzinatto2022dynamical,franzmann2017varying}) and in the data-fitting or observational approach (e.g. \cite{eaves2021constraints,gupta2020cosmology,gupta2021constraint,gupta2021testing,gupta2022effect,lee2021minimally,mendoncca2021search}).

In fact, recent studies of extensions of the general relativity theory propose that fundamental couplings of physics are expected to vary. As a few examples we mention models involving variations of the fine-structure constant $\alpha$~\cite{bora2021constraints,colacco2019galaxy,galli2013clusters,gonccalves2020variation,king2012spatial,kotuvs2017high,leefer2013new,liu2021probing,van2015variation}, models  admitting variation of the Newton's gravitational constant $G$
~\cite{dirac1937cosmological,garcia2011upper,jofre2006constraining,lazaridis2009generic,ooba2016planck,verbiest2008precision,vijaykumar2021constraints,zhao2018constraining}, and models addressing the variation of the speed of light $c$~\cite{FermiGBMLAT:2009nfe,Agrawal:2021cim,Albrecht:1998ir,Barrow:1999is,Cao:2016dgw,Cao:2018rzc,Cruz:2012bwp,liu2018light,Liu:2021eit,mendoncca2021search,Moffat:1992ud,Qi:2014zja,Salzano:2016hce,Salzano:2014lra,uzan2003fundamental,Xu:2016zxi,xu2016light,zhu2021pre}. The main feature of these works is the fact that they consider only one of the aforementioned couplings as a function of the time, while disregarding possible variations of the others. Models proposing variation of more than one coupling have been studied recently in Refs.
~\cite{bonanno2021effective,costa2019covariant,eaves2021constraints,franzmann2017varying,gupta2020cosmology,gupta2021constraint,gupta2021testing,Gupta:2020wgz,gupta2022faint,gupta2022effect,lee2021minimally}, which also include the cosmological constant $\Lambda$ as a new dynamic variable. 
Simultaneous variations of $\left\{ c,G,\Lambda\right\}$ were explored by Refs.~\cite{costa2019covariant,franzmann2017varying,gupta2020cosmology} in a framework called Co-varying Physical Couplings (CPC). The CPC framework is a modified gravity proposal inspired by Brans-Dicke
theory \cite{brans1961mach,faraoni2004cosmology} that can be understood as a generalization
to GR wherein the physical couplings are allowed
to co-vary. In a previous paper \cite{cuzinatto2022observational}, we have scrutinized the minimal CPC model, a model admitting co-varying $c$ and $G$ while $\Lambda$ is kept as a genuine constant. The minimal CPC model was constrained via gas mass fraction data---a powerful model agnostic observational probe \cite{Allen:2007ue,Allen:2011zs,Battaglia:2011cn,Ettori:2009wp,Holanda:2020sqm,Mantz:2014xba,Sarazin:1986zz}. The analysis therein favored no variation for both
$c$ and $G$. 

In the present paper we consider a full CPC model wherein the constancy of all the three fundamental couplings  $\{c,G, \Lambda\}$ is relaxed. This model shall be constrained via the $H(z)$ cosmic chronometers observational window \cite{Magana:2017nfs} and SNe Ia data \cite{Pan-STARRS1:2017jku}. In order to do so, we are led to review the important concepts of distances, luminosity and flux in the CPC framework, to deduce how they are modified in the presence of varying $\{c,G,\Lambda \}$, and to derive the generalized Hubble function in this context. The data fitting will indicated if the late-time cosmological data makes room for an entangled time variation of the trio $\{c,G,\Lambda \}$. 

The contents in the paper are organized as follows. In Section \ref{sec:CPC-Cosmo} we briefly review the CPC framework preparing the stage to introduce the
varying-$\{c,G,\Lambda\}$ model to be explored in the paper. We also specify CPC framework for cosmology in this section. Section \ref{sec:CPC-Observables} introduces the equations of CPC cosmology for contact with observation; this task includes the derivation of the equation ruling the entangled evolution of $c$, $G$ and $\Lambda$
, the calculation of the modified luminosity distance, the determination of the generalized distance modulus, and the specification of the normalized Hubble function within the CPC framework. Section \ref{sec:CPC-LinearLambda} finally presents the functional form of our full CPC model with $\Lambda=\Lambda(a)$, $c=c(a)$ and $G=G(a)$. It is shown that the CPC framework enables one to deduce the functional form of the functions $c=c(a)$ and $G=G(a)$ once the function $\Lambda=\Lambda(a)$ is provided. This deductive character is a key advantage of CPC framework: in this way, it keeps \emph{ad hoc} hypotheses to a bare minimum. Section \ref{sec:CPC-DataConstrained} is dedicated to constrain our co-varying $\{c,G,\Lambda\}$ model against observations. A discussion of the results is performed in Section \ref{sec:Discussion} where our conclusions are also presented.


\section{CPC framework in cosmology \label{sec:CPC-Cosmo}}



The CPC framework is based on the following assumptions: (i) gravity
is manifested as the dynamics of the metric tensor $g_{\mu\nu}$ equipping
a metric-compatible Riemannian four-dimensional manifold, i.e. $\nabla_{\rho}g_{\mu\nu}=0$;
(ii) matter and energy work as source of the gravitational field through
the energy momentum tensor $T_{\mu\nu}$ which is covariantly conserved,
i.e. $\nabla_{\mu}T^{\mu\nu}=0$; and (iii) the field equations for
$g_{\mu\nu}$ are formally the same as Einstein Field Equations of
GR but the couplings $\left\{ G,c,\Lambda\right\} $ therein are promoted
to functions of the spacetime coordinates, i.e. 
\begin{equation}
G_{\mu\nu}+\Lambda g_{\mu\nu}=\frac{8\pi G}{c^{4}}T_{\mu\nu}\label{eq:CPC-FE}
\end{equation}
are CPC Field Equations with $G=G\left(x^{\mu}\right)$, $c=c\left(x^{\mu}\right)$,
and $\Lambda=\Lambda\left(x^{\mu}\right)$. Eq. (\ref{eq:CPC-FE})
contains the regular Einstein tensor
\begin{equation}
G_{\mu\nu}=R_{\mu\nu}-\frac{1}{2}Rg_{\mu\nu},\label{eq:EinsteinTensor}
\end{equation}
which satisfies the Bianchi identity \cite{DeSabbata:1986sv}: $\nabla_{\mu}G^{\mu\nu}=0$.
The covariant derivative $\nabla=\partial+\Gamma$ is build with the
regular Christoffel symbols \cite{Carroll:2004st}.

By taking the covariant derivative of the field equation (\ref{eq:CPC-FE})
and using the features (i)--(iii) above, viz. $\nabla^{\mu}G_{\mu\nu}=\nabla^{\mu}g_{\mu\nu}=\nabla^{\mu}T_{\mu\nu}=0$,
we are led to 
\begin{equation}
\left[\frac{\partial_{\mu}G}{G}-4\frac{\partial_{\mu}c}{c}\right]\frac{8\pi G}{c^{4}}T^{\mu\nu}-\left(\partial_{\mu}\Lambda\right)g^{\mu\nu}=0,\label{eq:GC}
\end{equation}
which is an identity that must be satisfied for any consistent model
within the CPC scenario. Eq. (\ref{eq:GC}) is called the General
Constraint (GC).

Refs. \cite{costa2019covariant,cuzinatto2022observational,gupta2020cosmology} emphasized the consequences
of the GC for the modified gravity theory described by the CPC framework.
In fact, Eq. (\ref{eq:GC}) shows that eventual variations of the
couplings in the set $\left\{ G,c,\Lambda\right\} $ are coupled to
matter fields through the stress-energy tensor $T_{\mu\nu}$. Accordingly,
matter tells both spacetime how to curve and the couplings how to
run; conversely, spacetime and the co-varying couplings determine the
dynamics of the matter fields.

Another consequence that the GC entails is: the eventual variations
of the couplings $\left\{ G,c,\Lambda\right\} $ are not independent
(hence the term ``co-varying'' in the name of CPC framework). As an
example, one may consider the minimal model where $\Lambda$ is a
genuine constant. Then, $\partial_{\mu}\Lambda=0$ and Eq. (\ref{eq:GC})
yields
\begin{equation}
\frac{G}{G_{0}}=\left(\frac{c}{c_{0}}\right)^{4}\qquad\left(\Lambda=\text{const}\right),\label{eq:G(c)-minimal-CPC}
\end{equation}
for non-vacuum solutions. Eq. (\ref{eq:G(c)-minimal-CPC}) demonstrates
that a dynamical $c$ enforces a dynamical $G$ in a unequivocally determined
way.  This possibility was precisely the instance explored by \cite{cuzinatto2022observational}, where this minimal model was constrained
by means of the $f_{\text{gas}}$ observational window \cite{Allen:2007ue,Allen:2011zs,Battaglia:2011cn,Ettori:2009wp,Holanda:2020sqm,Mantz:2014xba,Sarazin:1986zz}.
As it happens, the data fitting disfavours any actual variations of
the speed of light $c$ and, consequently, of the gravitational coupling
$G$. In this paper we want to check if this result is confirmed by
other observational probes, such as cosmic chronometers---a.k.a $H(z)$
data \cite{Magana:2017nfs}---and SNe Ia data \cite{Pan-STARRS1:2017jku}. Perhaps
more important than that, in this paper we would like to verify if
a less restrictive model could favor co-varying physical couplings
within the CPC framework.

Accordingly, herein we propose a model where $\partial_{\mu}\Lambda\neq0$
and test it against the observations. Notice that a varying
$\Lambda$ will complicate the solution of the General Constraint
in Eq. (\ref{eq:GC}). In actuality, we will have to specify the matter-energy
content in order to resolve the GC and find the interdependence between
the couplings. For that, we need to choose the physical system under
scrutiny. If we are going to use cosmological data sets, we should
establish the key equations of CPC framework---e.g. Eqs. (\ref{eq:CPC-FE})
and (\ref{eq:GC})---in the context of background cosmology. This
is what we do next.

The standard assumption is that the universe is homogeneous and isotropic
in cosmological scales. According to this principle, the functions
$G$, $c$ and $\Lambda$ must depend on the time coordinate only.
Moreover, the metric tensor is read off the FLRW line element \cite{Weinberg:1972kfs}
\begin{equation}
ds^{2}=-c^{2}dt^{2}+a^{2}\left(t\right)\left[\frac{dr^{2}}{1-kr^{2}}+r^{2}\left(d\theta^{2}+\sin^{2}\theta d\varphi^{2}\right)\right],\label{eq:FLRW}
\end{equation}
where $k=-1,0,+1$ for a open, flat or closed spacial sector, respectively.
It should be kept in mind that $c=c\left(t\right)$ in (\ref{eq:FLRW})---and
everywhere else. In addition, the matter-energy content is modelled
by the stress-energy tensor of a perfect fluid \cite{Carroll:2004st}:
\begin{equation}
T_{\hphantom{\mu}\nu}^{\mu}=\text{diag}\left\{ -\varepsilon,p,p,p\right\}, \label{eq:T_munu-perfect-fluid}
\end{equation}
where $\varepsilon$ is the energy density and $p$ is the pressure. The time-dependent speed of light enters the energy density expression $\varepsilon = \rho c^2$, with $\rho$ denoting the (rest) matter density (for cases other than ultra-relativistic particles).
Substitution of (\ref{eq:FLRW}) and (\ref{eq:T_munu-perfect-fluid})
into (\ref{eq:CPC-FE}) leads to the Friedmann equations of CPC's
background cosmology \cite{costa2019covariant}: 
\begin{equation}
H^{2}=\frac{8\pi G}{3c^{2}}\varepsilon+\frac{\Lambda c^{2}}{3}-\frac{kc^{2}}{a^{2}},\label{eq:CPC-Friedmann-Eq}
\end{equation}
and
\begin{equation}
\frac{\ddot{a}}{a}=-\frac{4\pi G}{3c^{2}}\left(\varepsilon+3p\right)+\frac{\Lambda c^{2}}{3}+\frac{\dot{c}}{c}H,\label{eq:CPC-Acceleration-Eq}
\end{equation}
where dot denotes a time derivative, $H=\dot{a}/a$ is the Hubble
function, which is defined from the scale factor $a$ and its derivative.
Eq. (\ref{eq:CPC-Friedmann-Eq}) is formally the same as the first
Friedmann equation \cite{ellis2012relativistic,ryden2017introduction} but, obviously, $c=c\left(t\right)$,
$G=G(t)$ and $\Lambda=\Lambda\left(t\right)$ here. The second Friedmann
equation, or the acceleration equation, is explicitly modified
by the presence of the last term in Eq. (\ref{eq:CPC-Acceleration-Eq})
which depends on $\dot{c}$.

Since the energy momentum tensor is covariantly conserved within the
CPC scheme, it is not surprising that:
\begin{equation}
\dot{\varepsilon}+3\frac{\dot{a}}{a}\left(\varepsilon+p\right)=0,\label{eq:CPC-Continuity-Eq}
\end{equation}
i.e. the continuity equation holds. This not-so-strange (and even
natural) feature is not shared by all the models accommodating a varying
speed of light (VSL)---for a counter-example see e.g. Ref. \cite{Albrecht:1998ir}. 

Because $G$, $c$ and $\Lambda$ do not depend on the space coordinates
$\left(r,\theta,\varphi\right)$ in the context of background cosmology, Eq. (\ref{eq:GC}) reduces to: 
\begin{equation}
\left(\frac{\dot{G}}{G}-4\frac{\dot{c}}{c}\right)\frac{8\pi G}{c^{4}}\varepsilon+\dot{\Lambda}=0.\label{eq:CPC-GC-Cosmology}
\end{equation}
As a side note, the violation of the continuity equation (\ref{eq:CPC-Continuity-Eq})
observed in the VSL proposal by \cite{Albrecht:1998ir} equals precisely
the left hand side of our Eq.~(\ref{eq:CPC-GC-Cosmology})---see
Ref. \cite{costa2019covariant}. Consequently, the VSL scenario in \cite{Albrecht:1998ir}
does not respect the GC.

For solving the system composed by Eqs. (\ref{eq:CPC-Friedmann-Eq})--(\ref{eq:CPC-GC-Cosmology})
one should provide two constitutive equations. The first one is an
equation of state (EoS) relating the pressure to the energy density;
this is traditionally done through
\begin{equation}
p=w\varepsilon\qquad\left(w=\text{constant}\right).\label{eq:EOS}
\end{equation}
A dust-matter like component corresponds to $w=0$, while a radiation
content demands $w=1/3$. A constant cosmological constant can be
accounted for via the choice $w=-1$; here, however, $\Lambda$ is
kept explicit due to its fundamental role as a cosmological time-varying coupling. 

The second constitutive equation required in CPC cosmology is an ansatz for the time-varying function of one or more of the couplings in the
set $\left\{ G,c,\Lambda\right\} $. The ansatz(e) would have to be
consistent with the GC, Eq. (\ref{eq:CPC-GC-Cosmology}). For example,
the minimal CPC model with $\dot{\Lambda}=0$ and $G\propto c^{4}$
is consistent with
\begin{equation}
\frac{\dot{G}}{G}=4\frac{\dot{c}}{c}\qquad\left(\dot{\Lambda}=0\right). \label{eq:GdotOverG-minimalCPC}
\end{equation}
When $\Lambda$ is allowed to vary, Eq. (\ref{eq:CPC-GC-Cosmology})
will not automatically provide a relation between $G$ and $c$. This
difficulty can be resolved by the following idea. Inspired by the equation
above, Gupta proposed the following ansatz \cite{gupta2020cosmology}:
\begin{equation}
\frac{\dot{G}}{G}=\sigma\frac{\dot{c}}{c}\qquad\left(\dot{\Lambda}\neq0\right)\label{eq:GdotOverG-GuptaAnsatz}
\end{equation}
with $\sigma=\text{constant}$. Obviously, if $\Lambda=\text{constant}$
then $\sigma=4$ for compliance with the GC and Eq. (\ref{eq:GdotOverG-minimalCPC});
otherwise, the constant $\sigma$ is unconstrained in principle. In
the next sections we will see how Eq.~(\ref{eq:GdotOverG-GuptaAnsatz})
plays out in the context of a co-varying-$\Lambda$ model within the
CPC framework.


\section{Towards a varying-$\Lambda$ model in CPC cosmology}  \label{sec:CPC-Observables}

Let us prepare the equations of CPC cosmology for application to our model with varying $\Lambda$.


\subsection{Gupta's ansatz and the GC \label{sucsec:GC}}



Gupta's ansatz enables us to eliminate $G$ from the general constraint.
In fact, once $c$ is encountered then $G$ is automatically determined
by Eq. (\ref{eq:GdotOverG-GuptaAnsatz}). 

Continuity equation (\ref{eq:CPC-Continuity-Eq}) and equation of
state (\ref{eq:EOS}) yield:
\begin{equation}
\varepsilon=\varepsilon_{0}\left(\frac{a}{a_{0}}\right)^{-3\left(1+w\right)},\label{eq:epsilon(a)}
\end{equation}
just like in standard cosmology. As usual, $a_{0}=a\left(t_{0}\right)=1$
is the (normalized) value of the scale factor calculated at today's
time $t=t_{0}$. (Sometimes we will write $a_{0}$ explicitly to keep
track of the dimensions.) Similarly, $\varepsilon_0=\varepsilon(t_0)$ is the present-day value of energy density.

Let us introduce the following convenient parametrizations for $c=c\left(t\right)$
and $\Lambda=\Lambda\left(t\right)$:
\begin{equation}
c=c_{0}\phi_{c}\left(a\right),\qquad\text{and}\qquad\Lambda=\Lambda_{0}\phi_{\Lambda}\left(a\right),\label{eq:c(phi_c)_Lambda(phi_Lambda)}
\end{equation}
where $\phi_{c}=\phi_{c}\left(a\right)$ and $\phi_{\Lambda}=\phi_{\Lambda}\left(a\right)$
are dimensionless functions of the time-dependent scale factor $a=a\left(t\right)$.
Now the dimensions are carried exclusively by the present-day value
of the speed of light, $c_{0}=c\left(a_{0}\right)$, and today's
value of the cosmological term, $\Lambda_{0}=\Lambda\left(a_{0}\right)$.
Moreover, we take $\phi_{c,0}=\phi_{c}\left(a_{0}\right)=1$ and  $\phi_{\Lambda,0}=\phi_{\Lambda}\left(a_{0}\right)=1$. In this way we guarantee that it is the dimensionless part of the couplings  that actually vary. The necessity for the dimensionless feature in possibly varying couplings is strongly defended by some authors (e.g. \cite{Duff:2014mva,ellis2005c}).

Plugging Eqs. (\ref{eq:GdotOverG-GuptaAnsatz}), (\ref{eq:epsilon(a)})
and (\ref{eq:c(phi_c)_Lambda(phi_Lambda)}) into Eq. (\ref{eq:CPC-GC-Cosmology})
leads to the new form of the General Constraint:
\begin{equation}
\left(4-\sigma\right)\phi_{c}^{\left(\sigma-5\right)}\phi_{c}^{\prime}\,\Omega_{0}a^{-3\left(1+w\right)}=\Omega_{\Lambda,0}\phi_{\Lambda}^{\prime},\label{eq:CPC-GC-Gupta}
\end{equation}
where prime denotes differentiation with respect to the scale factor
$a$ (e.g. $\phi_{c}^{\prime}=\frac{d\phi_{c}}{da}$) and we have
made use of some of the energy density parameters' definitions \cite{ellis2012relativistic,ryden2017introduction}:
\begin{equation}
\Omega\left(a\right)\equiv\frac{\varepsilon\left(a\right)}{\varepsilon_{c,0}},\quad\varepsilon_{c,0}=\frac{3H_{0}^{2}c_{0}^{2}}{8\pi G_{0}},\quad\Omega_{\Lambda,0}=\frac{\Lambda_{0}c_{0}^{2}}{3H_{0}^{2}},\quad\Omega_{k,0}=-\frac{kc_{0}^{2}}{H_{0}^{2}a_{0}^{2}}.\label{eq:EnergyDensityParameters}
\end{equation}
Accordingly, $\Omega_0 = \Omega\left( a_0 \right) = \varepsilon_0 / \varepsilon_{c,0}$, with $\varepsilon_0 = \varepsilon \left( a_0 \right)$ representing the value of the energy density today.

Eq.~(\ref{eq:CPC-GC-Gupta}) is consistent with our previous comments.
If $\dot{\Lambda}=0$, then $\phi_{\Lambda}^{\prime}=0$ and either
$\sigma=4$ (as in the minimal CPC model where $G\propto c^{4}$)
or $\phi_{c}^{\prime}=0$ (which means a constant $c$, yielding a
constant $G$ due to the constraint in Gupta's ansatz: this is the
standard setting of non-varying fundamental constants).

Moreover, Eq.~(\ref{eq:CPC-Friedmann-Eq}) under Gupta's ansatz (\ref{eq:GdotOverG-GuptaAnsatz}) reads:
\begin{equation}
E^{2}\left(a\right)=\phi_{c}^{\left(\sigma-2\right)}\Omega\left(a\right)+\phi_{c}^{2}\phi_{\Lambda}\Omega_{\Lambda,0}+\left(\frac{\phi_{c}}{a}\right)^{2}\Omega_{k,0}\label{eq:CPC-Friedmann-Gupta}
\end{equation}
where $E\left(a\right)=\frac{H\left(a\right)}{H_{0}}$ is the normalized
Hubble function. While deriving Eq.~(\ref{eq:CPC-Friedmann-Gupta}),
we have utilized Eqs.~(\ref{eq:c(phi_c)_Lambda(phi_Lambda)}) and
(\ref{eq:EnergyDensityParameters}). Notice that at the present day $a=a_{0}$, $H\left(a_{0}\right)=H_{0}$, and consequently $E\left(a_{0}\right)=1$.
In this case, Eq.~(\ref{eq:CPC-Friedmann-Gupta}) gives---recall
that $\phi_{c,0}=\phi_{\Lambda,0}=1$:
\begin{equation}
1=\Omega_{0}+\Omega_{\Lambda,0}+\Omega_{k,0},\label{eq:CPC-Friedmann-today}
\end{equation}
which is the same result as in standard cosmology. 


\subsection{Contact with observations: distances, luminosity, and flux in the CPC framework \label{subsec:DistancesFluxLuminosity}}



The normalized Hubble function $E\left(a\right)$ is used for contact
with observations through the proper distance $d_{p}\left(a\right)$,
the measure of distance between the astrophysical source and the observer
in a dynamical spacetime in the presence of co-varying physical couplings
\cite{gupta2020cosmology}:
\begin{equation}
d_{p}\left(z\right)=\frac{c_{0}}{a_{0}H_{0}}\int_{0}^{z}\frac{1}{E\left(z\right)}\phi_{c}\left(z\right)dz.\label{eq:CPC-d_p}
\end{equation}
where $(1+z)=\left(a_{0}/a\right)$ defines the redshift, and $\phi_{c}=\phi_{c}\left(z\right)=\phi_{c}\left(a\left(z\right)\right)$
depends on the particular functional dependence of
the speed of light $c$ in terms of the scale factor $a$. The standard
result is recovered for $\phi_{c}=1$ which means $c=c_{0}=\text{constant}$,
cf. Eq. (\ref{eq:c(phi_c)_Lambda(phi_Lambda)}).

In practice, we measure magnitudes of the sources of photons. The
difference between the bolometric apparent magnitude $m$ and the
bolometric absolute magnitude $M$ defines the distance modulus $\mu$.
In fact \cite{ryden2017introduction}
\begin{equation}
\mu=m-M=5\log_{10}\left(\frac{d_{L}}{1\text{ Mpc}}\right)+25.\label{eq:mu(d_L)}
\end{equation}
This quantity depends on the luminosity distance $d_{L}$ which is
defined from the flux $F$ in an Euclidean static universe as 
\begin{equation}
F=\frac{L}{4\pi d_{L}^{2}}.\label{eq:f(d_L)}
\end{equation}
Here $L$ is the luminosity: energy given off by the source per unit
time. 

In the CPC framework the luminosity will be impacted by two different
effects, namely: (i) the regular redshift due to the Universe's expansion;
and (ii) the change underwent by the speed of light during this period. To make this
explicit, consider the photon energy $E$ (Planck's formula) as measured
by the observer at $x^{\mu}=\left(t_{0},0,0,0\right)$:
\begin{equation}
E_{0}=h\nu_{0}=\frac{hc_{0}}{\lambda_{0}}\Rightarrow E_{0}=\frac{\lambda_{e}}{\lambda_{0}}\left(\frac{hc_{e}}{\lambda_{e}}\right)\frac{c_{0}}{c_{e}}=\frac{\lambda_{e}}{\lambda_{0}}\frac{c_{0}}{c_{e}}E_{e},\label{eq:E_0(l,c,E_e)}
\end{equation}
where the quantities labelled with index $0$ ($e$) are measured
at the observers' (source's) position. $h$ is Planck's constant;
$\nu$ is the photon frequency and $\lambda$ is its wavelength.\footnote{It is important to emphasize that Eq.~(\ref{eq:E_0(l,c,E_e)}) would be different if one admits a varying Planck constant. That is precisely the case in Refs.~\cite{gupta2020cosmology,Gupta:2020wgz,gupta2022faint,gupta2022effect}. Thus it is not surprising that the conclusions in the present paper are different from the findings in the references just mentioned.} From
the definition of redshift as the fractional change in the radiation
wavelength, $z\equiv\left(\lambda_{0}-\lambda_{e}\right)/\lambda_{e}$,
one concludes that (\ref{eq:E_0(l,c,E_e)}) can be written as:
\begin{equation}
E_{0}=\frac{1}{\left(1+z\right)}\frac{c_{0}}{c_{e}}E_{e}.\label{eq:E_0(z,c)}
\end{equation}
The energy of the photon decreases as the universe expands: $E_{0}$
is smaller than $E_{e}$ for any $z>0$. On top of that, in CPC scenario,
the observed energy $E_{0}$ will change with respect to the emitted
energy $E_{e}$ according to the way the speed of light changes. For
a decreasing speed of light $c_{0}<c_{e}$ and the energy of the
photon decreases from emission to observation; for an increasing speed
of light $c_{0}>c_{e}$ and the photon energy increases from emission
to observation. So in a scenario of decreasing (increasing) speed
of light, the universe is younger (older) than in the standard picture
because the CMB photons cooled faster (slower) than what is regularly
expected.

What are the consequences of Eq.~(\ref{eq:E_0(z,c)}) upon the way we
measure luminosity and flux? In order to answer this question, let
$\delta t$ be the time interval between two wave crests (or two successive
photons emitted in the same direction). Then, in an expanding universe
\cite{gupta2020cosmology,ryden2017introduction}:
\begin{equation}
\frac{\lambda_{e}}{a_{e}}=\frac{\lambda_{0}}{a_{0}}\Rightarrow\frac{c_{e}\delta t_{e}}{a_{e}}=\frac{c_{0}\delta t_{0}}{a_{0}}.\label{eq:lambda_over_a}
\end{equation}
Notice that lengths are measured in terms of the speed of light: $\lambda=c\delta t$.
Due to the equation above, the time interval as measured by the observer
is:
\begin{equation}
\delta t_{0}=\frac{a_{0}}{a_{e}}\frac{c_{e}}{c_{0}}\delta t_{e}=\left(1+z\right)\frac{c_{e}}{c_{0}}\delta t_{e}.\label{eq:dt_0(z,c)}
\end{equation}
In our notation $a_{e}=a$ so that $\left(\frac{a_{0}}{a_{e}}\right)=1+z$.

We are now ready to decide how the luminosity changes in the CPC scenario.
Inserting (\ref{eq:E_0(z,c)}) and (\ref{eq:dt_0(z,c)}) into the
definition $L=E/\delta t$ leads to:
\begin{equation}
L_{0}=\frac{E_{0}}{\delta t_{0}}=\frac{1}{\left(1+z\right)^{2}}\left(\frac{c_{0}}{c_{e}}\right)^{2}L_{e}.\label{eq:CPC-L(z,c)}
\end{equation}
The term $\left(c_{0}/c_{e}\right)^{2}$ is exclusive of CPC scenarios. Eq.~(\ref{eq:CPC-L(z,c)}) reduces to the ordinary expression for the luminosity in the cosmological context if $c_e=c_0=$ constant.
We should point out that the Eq.~(\ref{eq:CPC-L(z,c)}) does not account
for correction on the calibration of SNe Ia peak luminosity curves,
corrections due to metallicity of the host galaxies, and other refinement
effects. Ref. \cite{gupta2022effect} discusses how these additional
effects could be accounted for and what would be the consequences
for constraining the cosmological parameter in the context of another
CPC model (with functional forms of $\left\{ G,c,\Lambda\right\} $
in terms of the redshift different from those explored here and the additional assumption that Planck's constant and Boltzmann's constant also vary).

The flux at the present day (at the observer's position) is:
\begin{equation}
F_{0}=\frac{L_{0}}{A_{p}\left(t_{0}\right)}=\frac{L_{0}}{L_{e}}\frac{L_{e}}{A_{p}\left(t_{0}\right)},\label{eq:f(L,A_p)}
\end{equation}
where $A_{p}\left(t_{0}\right)=4\pi d_{p}^{2}$ is the proper surface
area over which the emitted photons are spread out from the observer's
perspective. Substituting (\ref{eq:CPC-L(z,c)}) into (\ref{eq:f(L,A_p)})
to eliminate the ratio $\left(L_{0}/L_{e}\right)$ in terms of the
redshift and the speed of light gives:
\begin{equation}
F_{0}=\frac{L_{e}}{4\pi\left[d_{p}^{2}\left(\frac{c_{e}}{c_{0}}\right)^{2}\left(1+z\right)^{2}\right]}.\label{eq:f_0(z,c)}
\end{equation}
The comparison between (\ref{eq:f_0(z,c)}) and (\ref{eq:f(d_L)})---with
$F=F_{0}$ and $L=L_{e}$---enables us to recognize the expression
for the luminosity distance in CPC cosmology:
\begin{equation}
d_{L}\left(z\right)=\left(\frac{c_{e}}{c_{0}}\right)\left(1+z\right)d_{p}\left(z\right),\label{eq:CPC-d_L(z,c)}
\end{equation}
where $c_{e}=c\left(z\right)$. The usual expression is recovered if $\left( c_e/c_0 \right) = 1$. In the CPC framework, however, we generally have $\left( c_e/c_0 \right) = \phi_c(z)$, cf. Eq.~(\ref{eq:c(phi_c)_Lambda(phi_Lambda)}).

Plugging (\ref{eq:CPC-d_L(z,c)}) into (\ref{eq:mu(d_L)}) yields:
\begin{equation}
\mu=5\log_{10}\left(\frac{d_{p}\left(z\right)}{1\text{ Mpc}}\right)+5\log_{10}\left(1+z\right)+25 + 5 \log_{10}\left[\phi_{c}\left(z\right)\right].\label{eq:CPC-mu}
\end{equation}
The novelties due to CPC cosmology are the last term, which depends
on the varying speed of light explicitly, and in the argument of the
first term, through the $c$-dependent proper distance $d_{p}$---see
Eq. (\ref{eq:CPC-d_p}).


\section{Late-time CPC cosmology: a linear form for the varying-$\Lambda$ \label{sec:CPC-LinearLambda}}

In this section we introduce our particular co-varying-$\Lambda$ model within the CPC scheme.

The version of the General Constraint in Eq.~(\ref{eq:CPC-GC-Gupta})
encompasses two arbitrary functions $\phi_{c}\left(a\right)$ and $\phi_{\Lambda}\left(a\right)$.
Our strategy will be to propose a particular functional form of $\phi_{\Lambda}=\phi_{\Lambda}\left(a\right)$
and calculate $\phi_{c}=\phi_{c}\left(a\right)$ from (\ref{eq:CPC-GC-Gupta}). The working hypothesis is that the cosmological term $\Lambda$ is
approximately constant nowadays. This suggests a parametrization
of $\Lambda=\Lambda_{0}\phi_{\Lambda}\left(a\right)$ that is linear
in the scale factor:
\begin{equation}
\phi_{\Lambda}=\phi_{\Lambda,0}+\phi_{\Lambda,1}\left(1-\frac{a}{a_{0}}\right)=1+\phi_{\Lambda,1}\frac{z}{\left(1+z\right)},\label{eq:CPC-phi_Lambda}
\end{equation}
with $\left|\phi_{\Lambda,1}\right|\ll1$. Eq. (\ref{eq:CPC-phi_Lambda}) duly recovers $\phi_{\Lambda,0}=1$
for $a=a_{0}=1$ (and $z=0$). This form is similar to
an early parametrization of the EoS parameter $w$ in quintessence models (e.g.~\cite{amendola2010dark,Amendola:2006dg,Caldwell:1997ii}).

Apart from the $\Lambda$ contribution, that was separately taken into account in Eqs.~(\ref{eq:CPC-Friedmann-Eq}) and (\ref{eq:CPC-Acceleration-Eq}), the universe is vastly dominated by dust-like matter in the energy budget of the universe today \cite{ellis2012relativistic}. For this reason, we feel justified
to neglect the contribution of radiation and restrict our modelling of the matter-energy content in late-time
cosmology to the equation of state parameter value typical of pressureless
matter. Accordingly, we substitute $w=0$ and the ansatz (\ref{eq:CPC-phi_Lambda})
in Eq.~(\ref{eq:CPC-GC-Gupta}) for the GC. This procedure
leads to an equation for $\phi_{c}$, whose solution is:
\begin{align}
\phi_{c} & =\frac{\phi_{c,0}}{\left\{ 1-\frac{\Omega_{\Lambda,0}}{\Omega_{0}}\frac{1}{4}\frac{\phi_{\Lambda,1}}{\phi_{\Lambda,0}}\left[1-\left(\frac{a}{a_{0}}\right)^{4}\right]\right\} ^{1/\left(4-\sigma\right)}}\nonumber \\
 & =\left\{ 1-\frac{\left(1-\Omega_{m,0}\right)}{\Omega_{m,0}}\frac{\phi_{\Lambda,1}}{4}\left[1-\left(1+z\right)^{-4}\right]\right\} ^{-\frac{1}{\left(4-\sigma\right)}},\label{eq:CPC-phi_c}
\end{align}
where 
\begin{equation}
\Omega_{0}=\Omega_{m,0},\qquad\Omega_{k,0}=0,\qquad\Omega_{\Lambda,0}=\left(1-\Omega_{m,0}\right),\label{eq:CPC-LinearLambda-Omegas}
\end{equation}
i.e., radiation is negligible, we assume a flat universe, and we use
Eq.~(\ref{eq:CPC-Friedmann-today}) to write $\Omega_{\Lambda,0}$
in terms of the dust-like matter energy density parameter (in today's
value). Recall that $\phi_{c,0}=1$.

Notice that Eq.~(\ref{eq:CPC-phi_c}) requires $\sigma\neq4$, i.e.
it is adequate to describing a varying-$\Lambda$ scenario within CPC
(but it is not consistent with a constant $\Lambda$, such as the
minimal CPC model \cite{cuzinatto2022observational}). Moreover, Eq.~(\ref{eq:CPC-phi_c})
implies that the speed of light scales as $c\sim c_{0}a^{-4/\left(4-\sigma\right)}$.
A decreasing speed of light is consistent with the general prejudice
in the some works on VSL models involving phase-transition scenarios
\cite{Albrecht:1998ir,Barrow:1999is,Petit:1988ti,Petit:1995ys,Petit:2008eb,Shojaie:2004xw,Shojaie:2004sq}, so it might be that $0<\sigma<4$.
This interval will be used to reduce the space of parameters allowed
for $\sigma$ when we fit our varying-$\Lambda$ model to observational
data in the next section. 

Eqs.~(\ref{eq:CPC-phi_Lambda}) and (\ref{eq:CPC-phi_c}) constitute
the CPC model we call ``Co-varying-$\Lambda$ Dark Energy''. It is
``co-varying'' because all the couplings $\left\{ G,c,\Lambda\right\} $
vary together in this model. In fact, Eqs.~(\ref{eq:c(phi_c)_Lambda(phi_Lambda)}),
(\ref{eq:CPC-phi_Lambda}), and (\ref{eq:CPC-phi_c}) determine $c\left(z\right)=c_{0}\phi_{c}\left(z\right)$
and $\Lambda\left(z\right)=\Lambda_{0}\phi_{\Lambda}\left(z\right)$;
then, substitution of (\ref{eq:CPC-phi_c}) into (\ref{eq:GdotOverG-GuptaAnsatz})
specifies to $G\left(z\right)=G_{0}\phi_{c}^{\sigma}$. We also use
the soubriquet ``dark energy'' since the benchmark model of cosmology---the
$\Lambda$CDM model---explains the current cosmic acceleration, and
evidence of dark energy, through $\Lambda$ \cite{amendola2010dark}.


\section{Constraining the model from observational data \label{sec:CPC-DataConstrained}}

In order to constrain the free parameters of the linearly varying-$\Lambda$ model within CPC framework by using $H\left(z\right)$ and
SNe Ia data sets, the essential equation is the normalized Hubble function (\ref{eq:CPC-Friedmann-Gupta}), which in a flat background can be written as:
\begin{align}
& E^{2}\left(z\right) =\frac{H^{2}(z)}{H_{0}^{2}}\nonumber \\
& =\left[\phi_{c}\left(z\right)\right]^{\left(\sigma-2\right)}\Omega_{m,0}\left(1+z\right)^{3}+\left(1-\Omega_{m,0}\right)\phi_{\Lambda}\left(z\right)\left[\phi_{c}\left(z\right)\right]^{2},\label{eq:CPC-E(z)}
\end{align}
where it was used $\Omega\left(a\right)=\Omega_{m}=\Omega_{m,0}\left(1+z\right)^{3}$ for $w=0$ from Eq. (\ref{eq:epsilon(a)}). Additionally, the expressions  (\ref{eq:CPC-d_p}) for $d_{p}\left(z\right)$ and (\ref{eq:CPC-mu}) for $\mu$,
both adapted for the CPC scenario, complete the set of equations needed to constrain the model.

For the Hubble parameter data we use 31 differential age $H(z)$ data taken from \cite{Magana:2017nfs}, known as Cosmic Chronometers. This sample is model independent, covering the redshift range $0.07 < z < 1.965$. For the Supernovae type Ia data set we consider 1048 data from the Pantheon sample \cite{Pan-STARRS1:2017jku}. It is one of the largest combined sample of Supernovae, with redshift in the range $0.01 < z < 2.3$. 

The best-fit values and uncertainty of the parameters are obtained by maximizing the likelihood distribution
function of the combined set of data, which must be proportional to $e^{-\frac{1}{2}(\chi^2_{H}+\chi^2_{\mathrm{SN}})}$ where:
\begin{equation}
\chi^2_H = \sum_{i = 1}^{31}\frac{{\left[ H_{\mathrm{obs},i} - H(z_i,\mathbf{p})\right] }^{2}}{\sigma^{2}_{H_\mathrm{obs},i}},
\label{chi2H}
\end{equation}
is the $\chi^2$ function for $H(z)$ data, with $\mathbf{p}$ the vector of free parameters of the model, and
\begin{equation}
\chi^2_\mathrm{SN}=S_{mm}-\frac{S_m^2}{S_A},
\end{equation}
where $S_m=\sum_{i,j}\Delta m_i(C^{-1})_{ij}=\mathbf{\Delta m}^T\cdot\mathbf{C}^{-1}\cdot\mathbf{1}$, $S_{mm}=\sum_{i,j}\Delta m_i\Delta m_j(C^{-1})_{ij}=\mathbf{\Delta m}^T\cdot\mathbf{C}^{-1}\cdot\mathbf{\Delta m}$ and $S_A=\sum_{i,j}(C^{-1})_{ij}=\mathbf{1}^T\cdot\mathbf{C}^{-1}\cdot\mathbf{1}$. $\textbf{C}$ is a covariance matrix including statistical and systematic uncertainties \cite{SDSS:2014iwm}, and  $\mathbf{\Delta m}=m_{\mathrm{obs}}-\mu$, where $m_{\mathrm{obs}}$ is the observed magnitude and $\mu$ is given by Eq.~(\ref{eq:CPC-mu}).

By using the Affine Invariant method of Monte Carlo Markov Chain (MCMC) analysis implemented in {\sffamily Python} language with {\sffamily emcee} software \cite{Foreman-Mackey:2012any,goodman2010communications}, the constraints over the free parameters of the model are obtained though sampling the combined likelihood function.

The scenario described by Eq.~(\ref{eq:CPC-E(z)}) will be analyzed in the light of $H(z)$ data and SN data for two specific models. The first model is the minimal CPC model; it is dubbed ``Case I'' and appears in Section \ref{subsec:minimal-CPC-model}. The minimal CPC model admits the co-varying pair $\{G,c\}$ while $\Lambda$ is kept as a genuine constant. This model was explored in Ref.~\cite{cuzinatto2022observational} where it was constrained via $f$-gas data; the analysis therein favored no variation for both $G$ and $c$. Herein, the minimal CPC model is considered as a control model: we want to check if $H(z)$ data and SN Ia data also indicate that the pair $\{G,c\}$ should be constant. 

The second model to be scrutinized here is the varying-$\{G,c,\Lambda\}$ model, dubbed  Co-varying-$\Lambda$ Dark Energy, introduced in Section \ref{sec:CPC-LinearLambda}, and constrained in the upcoming Section \ref{subsec:LinearLambda-CPC-Model}. We call ``Case II'' the instance where the data fitting assumes no prior over the value of $H_0$; this analysis is performed in Sub-section \ref{subsubsec:LinearLambda-NoPrior}. Conversely, ``Case III'' is the name of the analysis where the model was fit to SNe Ia data and $H(z)$ data by assuming a prior over $H_0$; the details of this process are given in Sub-section \ref{subsubsec:LinearLambda-WithPrior}.


\subsection{Case I -- Minimal CPC model ($\dot{\Lambda}=0$, $\sigma = 4$) \label{subsec:minimal-CPC-model}}

In the first approach, we consider the case of a minimal CPC model, where $\sigma = 4$ and $\phi_\Lambda =1$ in Eq.~(\ref{eq:CPC-E(z)}). In this case,  $\{c,\,G \}$ are the only varying parameters, $\Lambda$ is constant, and Eqs.~(\ref{eq:G(c)-minimal-CPC}) and (\ref{eq:GdotOverG-minimalCPC}) are satisfied. We study two different parametrizations (P) for $c(z)$:
\begin{equation}
c\left(z\right)=c_{0}\phi_{c}\left(z\right)\,,\qquad\phi_{c}\left(z\right)\equiv\begin{cases}
1 + c_{1} z & \quad \left(\text{P I}\right)\\
1 + c_{1} \frac{z}{\left(1+z\right)} & \quad \left(\text{P II}\right)
\end{cases},
\label{eq:c(z)}
\end{equation}
where $c_1$  represents the deviation from the
constant value $c_{0}=299\,792\,458\,\text{m/s}$.\footnote{Parametrizations (P I) and (P II) in Eq.~(\ref{eq:c(z)}) are similar to the functional form adopted for $w(z)$ in some dynamical dark energy (DDE) models---see e.g. Table I in Ref.~\cite{Colgain:2021pmf}.}


\begin{figure}[h!]
\centering
\includegraphics[width=0.5\linewidth]{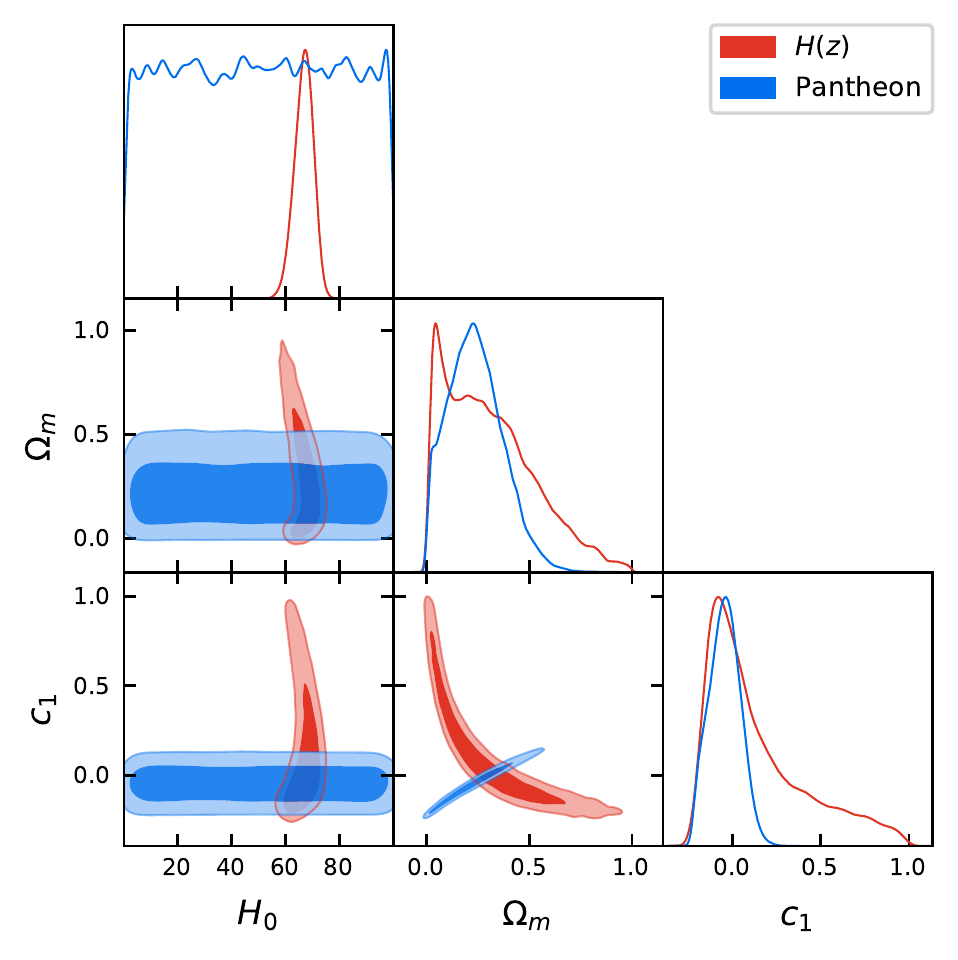}
\caption{
Constraints on the parameters set $\{H_0, \Omega_m,c_1\}$ of the minimal CPC model (co-varying-$\{c,G\}$ and $\Lambda =$ constant), \textbf{Case I} with parametrization \textbf{P I}. The panels display statistical contours at 68\% and 95\% c.l. from fitting to $H(z)$ data (red) and to SNe Ia-Pantheon data (blue).}
\label{c11fig01}
\end{figure}

\begin{figure}[h!]
\centering
\includegraphics[width=0.5\linewidth]{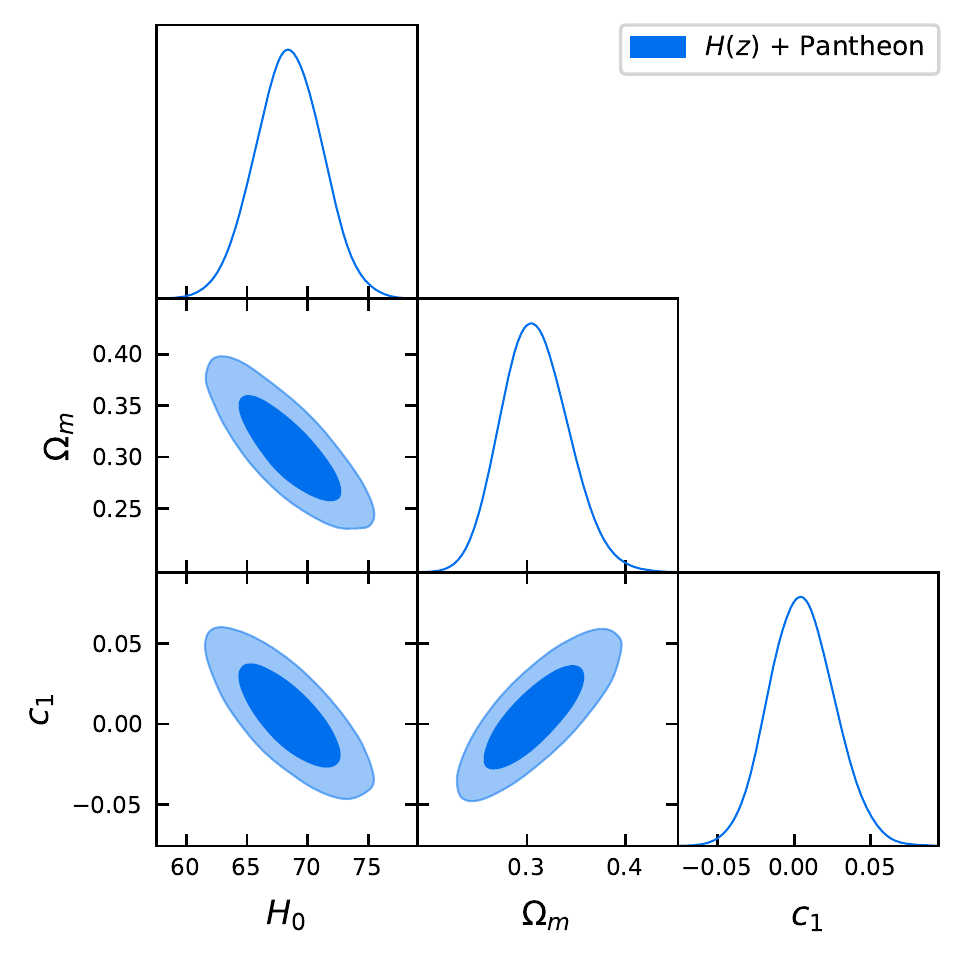}
\caption{Contours at 68\% and 95\% c.l. from the joint analysis of $H(z) \, + $ SNe Ia-Pantheon data sets for the parameters  $H_0$, $\Omega_m$ and $c_1$ of the minimal CPC model (co-varying-$\{c,G\}$ and $\Lambda =$ constant), \textbf{Case I} with parametrization \textbf{P I}.}
\label{c11fig02}
\end{figure}

\begin{figure}[h!]
\centering
\includegraphics[width=0.5\linewidth]{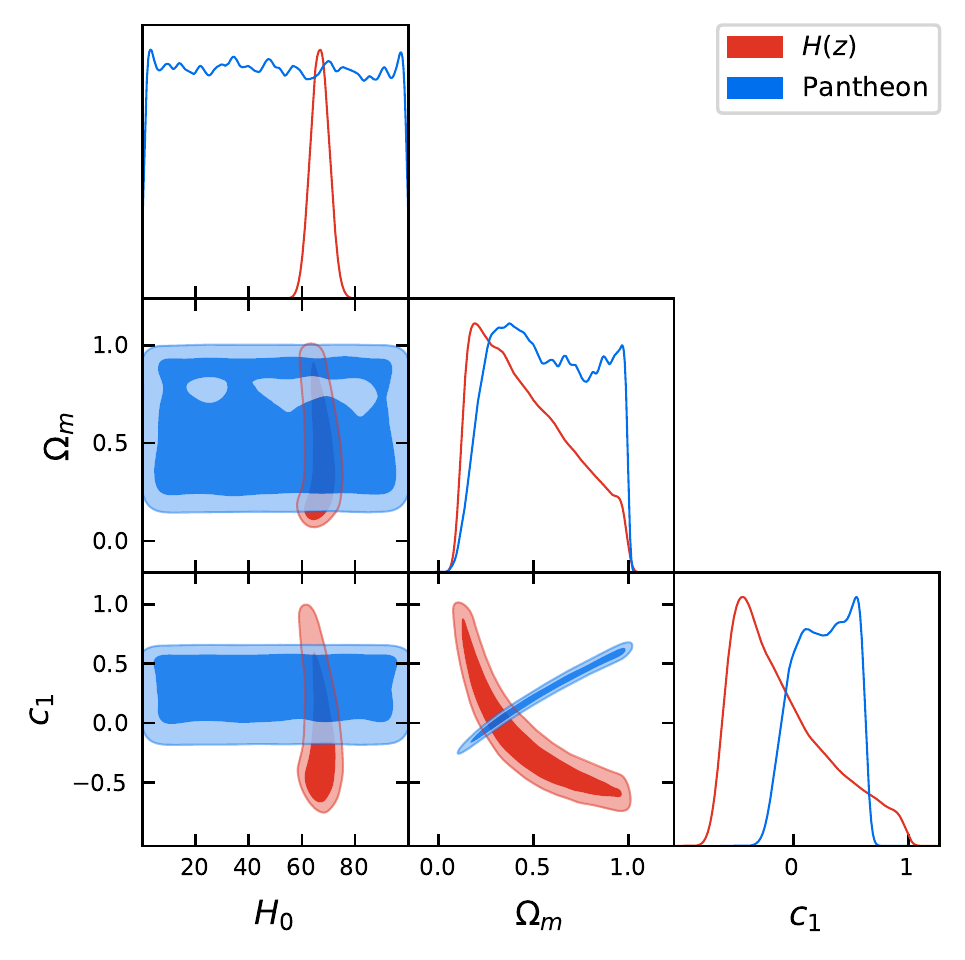}
\caption{
Constraints on the parameters set $\{H_0, \Omega_m,c_1\}$ of the minimal CPC model (co-varying-$\{c,G\}$ and $\Lambda =$ constant), \textbf{Case I} with parametrization \textbf{P II}. The panels display statistical contours at 68\% and 95\% c.l. from fitting to $H(z)$ data (red) and to SNe Ia-Pantheon data (blue).}
\label{c12fig01}
\end{figure}
\begin{figure}[h!]
\centering
\includegraphics[width=0.5\linewidth]{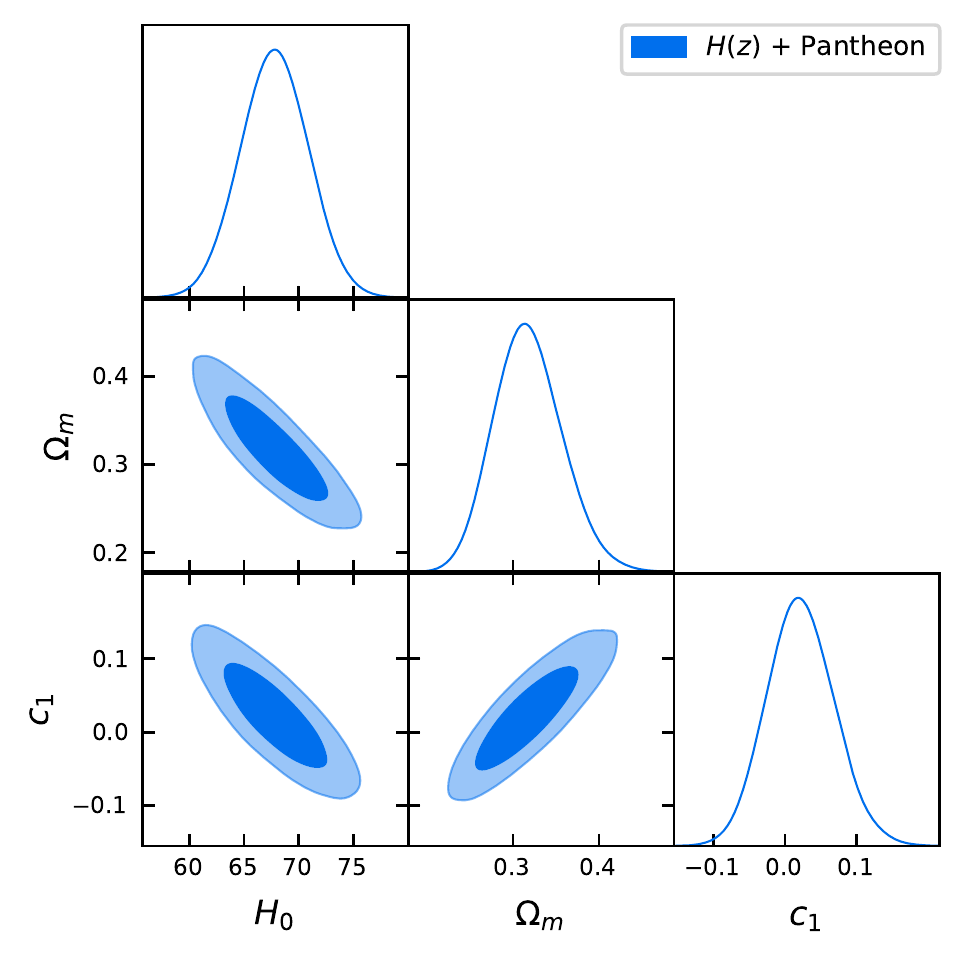}
\caption{Contours at 68\% and 95\% c.l. from the joint analysis of $H(z) \, + $ SNe Ia-Pantheon data sets for the parameters  $H_0$, $\Omega_m$ and $c_1$ of the minimal CPC model (co-varying-$\{c,G\}$ and $\Lambda =$ constant), \textbf{Case I} with parametrization \textbf{P II}.}
\label{c12fig02}
\end{figure}

The parameters to be constrained here are $[H_0,\,\Omega_m,\,c_1]$.
Figures \ref{c11fig01} and \ref{c11fig02} show the contours at 68- and 95-per cent confidence level (c.l.) for the separate sets of data, namely $H(z)$ data (red) and SNe Ia--Pantheon data (blue), and for the joint analysis, respectively, for P I. Figures \ref{c12fig01} and \ref{c12fig02} show the contours at 68- and 95-per cent confidence level (c.l.) for the separate sets of data and for the joint analysis, respectively, for P II. We see that, for both P I and P II, $H(z)$ data do not constrain the parameter $c_1$ very well while Supernovae data do not constrain the $H_0$ value efficiently. However the joint analysis is able to bring interesting constraints on those parameters. The mean value of the parameters and 68-, 95-, and 99-per cent c.l. limits are presented in Table \ref{tab01}. It is clear from it that the values of $H_0$ and $\Omega_m$ for both parametrizations are in good agreement with latest Planck 2018 \cite{Planck:2018vyg} results, and the values for the correction $c_1$ are compatible with zero within $1\sigma$ c.l. in both cases. This points to a non-variation of the speed of light, at least in this scenario of the minimal CPC model. This result agrees with the recent conclusion presented in Ref. \cite{cuzinatto2022observational}, which considered the same parametrizations as in our Eq.~(\ref{eq:c(z)}) and used Gas Mass Fraction data to constraint the parameter $c_1$.


\bigskip{}


\begin{table}[h!]
    \centering
     
    \begin{tabular} { l  c  c}

 Parameter &  P I  &  P II\\
\hline
{\boldmath$H_0            $} & $68.4^{+2.8 \, +5.5\, +8}_{-2.8\, -5.6\, -8}      $  &   $67.8^{+3.1 \, +6.2\, +9}_{-3.1\, -6.2\, -9}      $ \\


{\boldmath$\Omega_{m}         $} & $0.309^{+0.031\, +0.073\, +0.11}_{-0.036\, -0.064\, -0.090}   $ & $0.318^{+0.036\, +0.084\, +0.13}_{-0.043\, -0.074\, -0.11}   $ \\


{\boldmath$c_1         $} & $0.005^{+0.021\, +0.043\, +0.067}_{-0.021\, -0.042\, -0.061}   $ & $0.022^{+0.047\, +0.098\, +0.15}_{-0.047\, -0.091\, -0.14}   $\\

\hline
\end{tabular}
    \caption{Mean values of the parameters $H_0 (\mathrm{km}\,\mathrm{s}^{-1}\mathrm{Mpc}^{-1})$, $\Omega_m$ and $c_1$ and 68\%, 95\% and 99\% c.l. limits for Case I, parametrizations P I e P II.}
    \label{tab01}
\end{table}


\newpage


\subsection{Co-varying-$\Lambda$ Dark Energy ($\dot{\Lambda}\neq 0$)} \label{subsec:LinearLambda-CPC-Model}

In the second approach we consider the full CPC model where $c$, $G$ and $\Lambda$ vary simultaneously. In doing so, the set $\{c,\,G,\,\Lambda \}$ satisfies Eq.~(\ref{eq:GdotOverG-GuptaAnsatz}) with $\Lambda(z)$ and $c(z)$ given by Eqs.~(\ref{eq:CPC-phi_Lambda}) and (\ref{eq:CPC-phi_c}), respectively.


\subsubsection{Case II -- $\dot{\Lambda}\neq 0$ no prior over $H_0$ } \label{subsubsec:LinearLambda-NoPrior}

Let us first constrain the model without forcing any prior over the value of $H_0$. We will perform two different analysis. The first one maintains $\sigma$ as an open free parameter within the interval $[-4,+4]$ and the second analysis sets $\sigma = 3$.


\paragraph{(A) Keeping $\sigma$ within the interval $[-4,+4]$}

Following the discussion in the paragraph below Eq.~(\ref{eq:CPC-LinearLambda-Omegas}) in Section \ref{sec:CPC-LinearLambda}, we let $\sigma$ vary in the interval $-4.0 < \sigma < +4.0$, as an initial prior. This is done in order to assess if there is any particular value of $\sigma$ that is favored by this kind of data set. In fact, Refs. \cite{cuzinatto2022dynamical,eaves2021constraints,gupta2020cosmology,gupta2021constraint,gupta2021testing,Gupta:2020wgz,gupta2022faint,gupta2022varying,gupta2022effect} among others indicate that it should be $\sigma=3$. It would be very compelling indeed if this fact is verified within our Co-varying-$\Lambda$ Dark Energy model through data fitting.  

Here, we also use the prior $-2.0 < \phi_{\Lambda, 1}< +2.0$. This interval certainly includes the regime $\left| \phi_{\Lambda,1} \right| \ll 1 $ assumed by our linearly varying $\Lambda(a)$ below Eq.~(\ref{eq:CPC-phi_Lambda}).

The set of parameters to be constrained is $[H_0,\Omega_m,\phi_{\Lambda,1},\sigma]$. $H(z)$ data and SNe Ia-Pantheon data can be used to build the statistical contours at 68\% and 95\% c.l. for this set of parameters. The analysis reveals that the observational windows utilized here is unable to determine any preferred value for $\sigma$. Moreover, $H(z)$ data do not constrain $\phi_{\Lambda,1}$ while Supernovae data is only able to put a lower limit to the possible values of $\phi_{\Lambda,1}$. For these reasons, the present analysis (letting $\sigma$ free withing the interval $[-4,+4]$) is not very informative. Unfortunately, the data fitting with $H(z)$ and SNe Ia data does not favor $\sigma=3$ in the context of our particular CPC model as it would be so compelling.


\paragraph{(B) Setting $\sigma=3$}

Here we fix $\sigma = 3$ in order to extract the maximum amount of information about the new parameter $\phi_{\Lambda, 1}$ of our model via data fitting. The parameters to be constrained are $[H_0,\,\Omega_m,\,\phi_{\Lambda, 1}]$.

\begin{figure}[h!]
\centering
\includegraphics[width=0.5\linewidth]{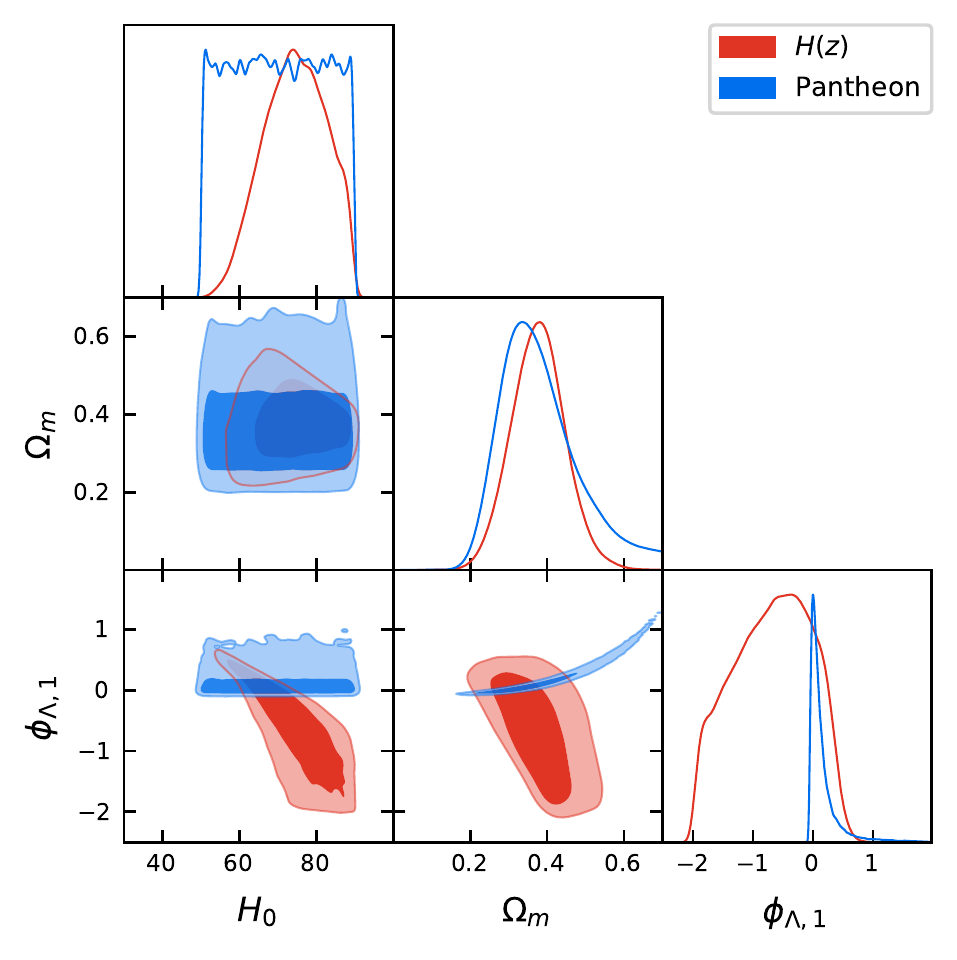}
\caption{
Constraints on the parameters set $\{H_0, \Omega_m,\phi_{\Lambda,1}\}$ of the full CPC model (simultaneously varying-$\{c,G,\Lambda\}$), \textbf{Case II} with $\sigma = 3$. The panels display statistical contours at 68\% and 95\% c.l. from fitting to $H(z)$ data (red) and to SNe Ia-Pantheon data (blue).}
\label{c22fig01}
\end{figure}

Figure \ref{c22fig01} shows the contour plots at 68\% and 95\% c.l. for the separate data sets, namely $H(z)$ data (red) and SNe Ia - Pantheon (blue), for the parameters  $H_0$ $\Omega_m$ and $\phi_{\Lambda,1}$. From the figure it is clear that $H(z)$ data do not constrain the parameter $\phi_{\Lambda,1}$ effectively. On the other hand, Supernovae data is not able to constrain the $H_0$ value efficaciously.

\begin{figure}[h!]
\centering
\includegraphics[width=0.5\linewidth]{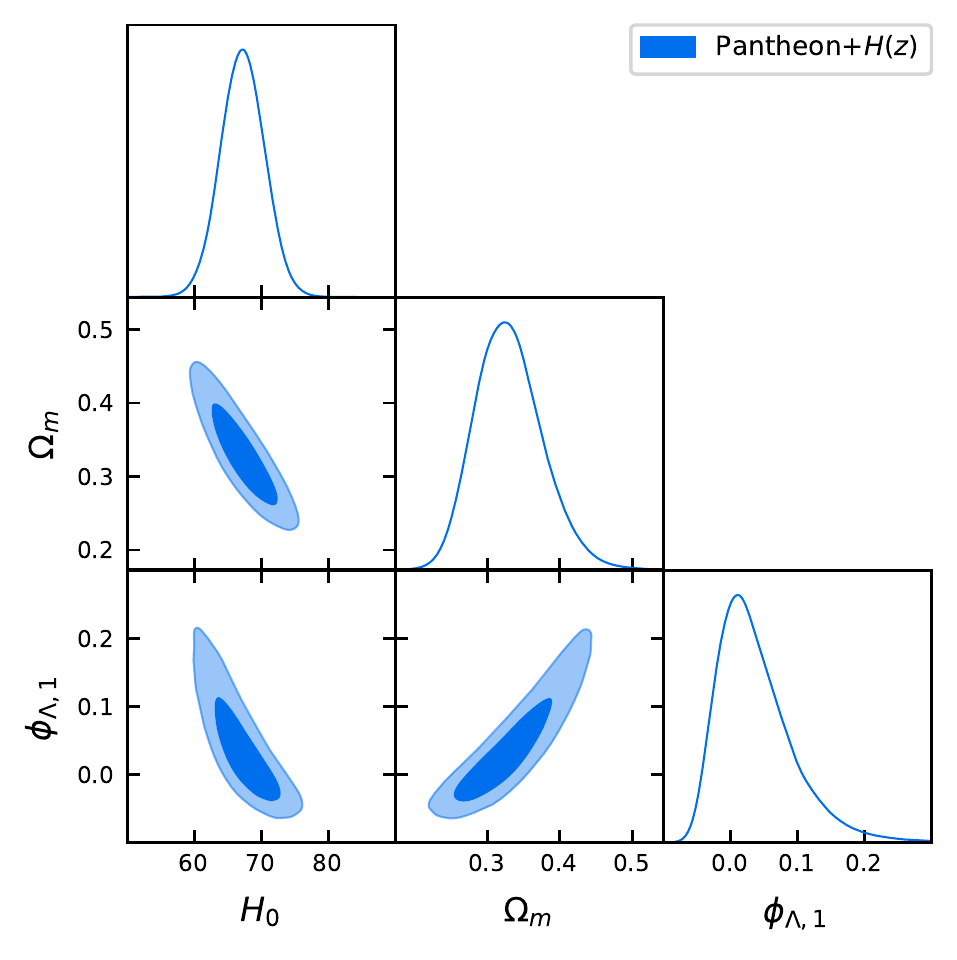}
\caption{Contours at 68\% and 95\% c.l. for the joint analysis of $H(z) \, +$ SNe Ia-Pantheon for the parameters  $H_0$, $\Omega_m$ and $\phi_{\Lambda,1}$ of the full CPC model (co-varying-$\{c,G,\Lambda\}$), \textbf{Case II} with $\sigma=3$.}
\label{c22fig02}
\end{figure}

A joint analysis of $H(z)$ data and SNe Ia data is displayed on Figure \ref{c22fig02}. The mean value of the parameters and 68\%, 95\% and 99\% c.l. limits are presented in Table \ref{tab02}. We conclude that the values of $H_0$ and $\Omega_m$ are in full agreement to the latest Planck 2018 \cite{Planck:2018vyg} results. The value for the correction $\phi_{\Lambda,1}$ is compatible with zero within $1\sigma$ c.l., which points toward a non-variation of the cosmological constant. This finding is true at least within our linearly co-varying-$\Lambda$ model with fixed $\sigma = 3$. 

\bigskip{}


\begin{table}[h!]
    \centering
    \begin{tabular} { l  c}

 Parameter &  Mean values\\
\hline
{\boldmath$H_0            $} & $67.2^{+3.3 \, +6.5\, +10}_{-3.3\, -6.5\, -10}      $\\


{\boldmath$\Omega_{m}         $} & $0.330^{+0.039\, +0.095\, +0.17}_{-0.052\, -0.088\, -0.12}   $\\


{\boldmath$\phi_{\Lambda,1}         $} & $0.039^{+0.031\, +0.12\, +0.31}_{-0.068\, -0.095\, -0.11}   $\\

\hline
\end{tabular}
    \caption{Mean values of the parameters $H_0 (\mathrm{km}\,\mathrm{s}^{-1}\mathrm{Mpc}^{-1})$, $\Omega_m$ and $\phi_{\Lambda,1}$ and 68\%, 95\% and 99\% c.l. limits for Case II, $\sigma=3$.}
    \label{tab02}
\end{table}


\bigskip{}



\newpage


\subsubsection{Case III -- $\dot{\Lambda}\neq 0$ with prior over $H_0$} \label{subsubsec:LinearLambda-WithPrior}

In this section's analysis, we test the same model as in Case II (B), i.e. with $\sigma=3$, but now impose current priors over the Hubble constant, which is quite constrained nowadays. Due to the so-called $H_0$ tension \cite{di2021realm}, we choose to perform two separate analysis: the first one uses the prior over $H_0$ from CMB data, cf. Planck18 \cite{Planck:2018vyg}, and the second analysis utilizes the prior over $H_0$ from Cepheids $+$ SNe Ia data, cf. \cite{Riess:2020fzl}.


\paragraph{(A) Planck prior over $H_0$}

Here we have used a Gaussian prior over $H_0$ from Planck 2018 CMB analysis, namely $H_0=67.36\pm0.54$ km s$^{-1}$ Mpc$^{-1}$. We have combined this prior with the $H(z)$ data from 31 cosmic chronometers \cite{Magana:2017nfs} and Pantheon SNe Ia data \cite{Pan-STARRS1:2017jku}. The results can be seen on Table \ref{tab03}.

\bigskip{}
%


\begin{table}[h!]
    \centering
    \begin{tabular} { l  c}
 Parameter &  Mean values\\
\hline
{\boldmath$H_0            $} & $67.37^{+0.53+1.1+1.6}_{-0.53-1.0-1.6}$\\

{\boldmath$\Omega_{m}     $} & $0.326^{+0.023+0.048+0.073}_{-0.023-0.046-0.068}   $\\

{\boldmath$\phi_{\Lambda, 1}$} & $0.027^{+0.025+0.059+0.10}_{-0.031-0.055-0.073}   $\\
\hline
\end{tabular}
    \caption{Mean values of the parameters $H_0$ (km s$^{-1}$Mpc$^{-1}$), $\Omega_m$ and $\phi_{\Lambda,1}$ and 68.3\%, 95.4\% and 99.7\% c.l. limits for Case III, $\sigma=3$, with Planck prior over $H_0$.}
    \label{tab03}
\end{table}


%
\bigskip{}

As can it be seen from this Table \ref{tab03}, $H_0$ is mainly constrained from Planck data to assume the value $H_0=67.37\pm0.53$ km s$^{-1}$ Mpc$^{-1}$ at 1$\sigma$ c.l. The result 
$\phi_{\Lambda,1}=0.027^{+0.025}_{-0.031}$ at 1$\sigma$ c.l., shows that the $\Lambda$ constancy condition, $\phi_{\Lambda,1}=0$, can not be discarded by this analysis.

\begin{figure}[h!]
    \centering
    \includegraphics[width=0.5\linewidth]{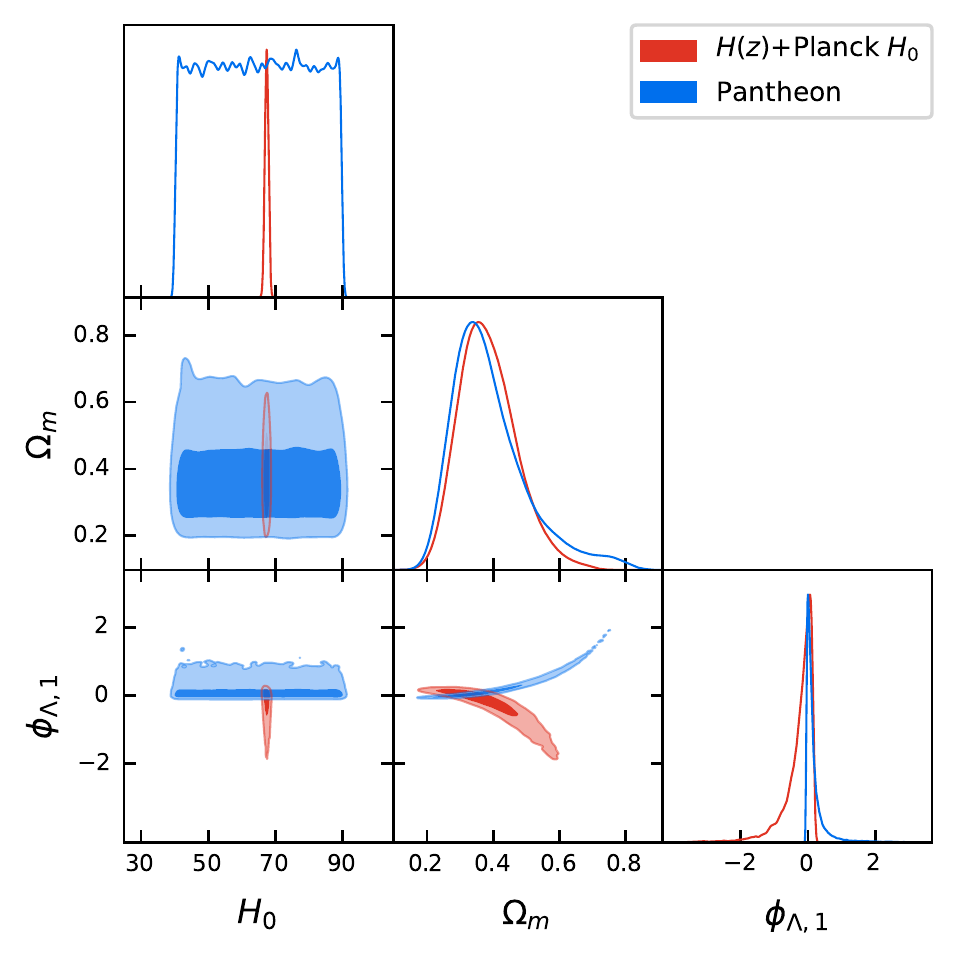}
    \caption{Constraints on parameters $\{H_0,\Omega_{m,0},\phi_{\Lambda,1}\}$ in \textbf{Case III (A)} of the  co-varying-$\{c,G,\Lambda\}$ model from two sets of data: (i) $H(z)$ data + $H_0$ prior from Planck data (red); and (ii) Pantheon data (blue). The statistical contours show 68.3\% and 95.4\% c.l. regions. }
    \label{HzPlanckH0}
\end{figure}

\begin{figure}[h!]
    \centering
    \includegraphics[width=0.5\linewidth]{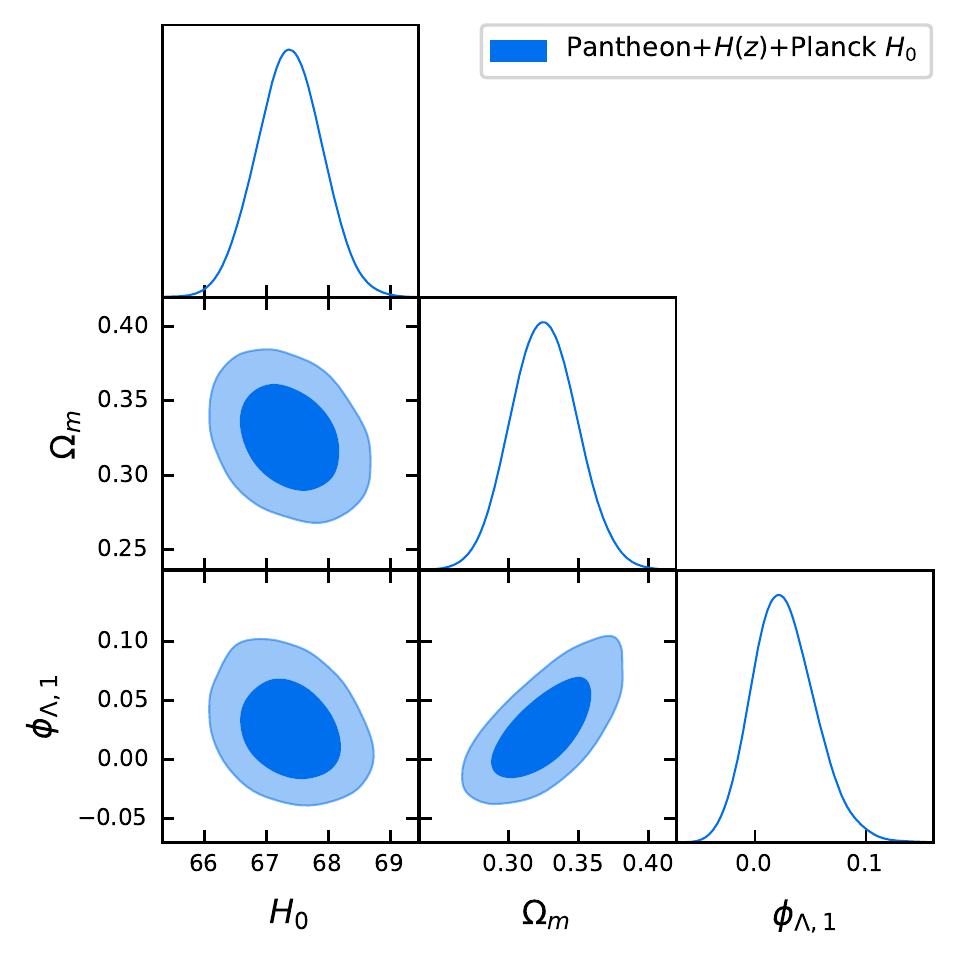}
    \caption{Constraints on the parameters $\{H_0,\Omega_{m,0},\phi_{\Lambda,1}\}$ in \textbf{Case III (A)} of the co-varying-$\{c,G,\Lambda\}$ model from $H(z)$ data + Planck $H_0$ + Pantheon, showing 68.3\% and 95.4\% c.l. statistical contours.}
    \label{HzPlanckH0comb}
\end{figure}

Fig. \ref{HzPlanckH0} shows the analysis of $H(z)$ data plus $H_0$ prior from Planck data and the separate analysis with SNe Ia data. Fig. \ref{HzPlanckH0comb} shows the contours for the joint analysis, which confirms the results shown in Table \ref{tab03}.

\bigskip{}
%


%
\bigskip{}
%

\newpage


\paragraph{(B) SH0ES prior over $H_0$}

In this analysis we utilize the local constraint from SH0ES over $H_0$. Accordingly, we admit a Gaussian prior $H_0=73.2\pm1.3$ km s$^{-1}$ Mpc$^{-1}$. We have combined this prior with $H(z)$ data and SNe Ia data. The results can be seen on Table \ref{tab04}.


\begin{table}[h!]
    \centering
    \begin{tabular} { l  c}
 Parameter &  Mean values\\
\hline
{\boldmath$H_0            $} & $72.4^{+1.2+2.4+3.6}_{-1.2-2.4-3.5}$\\

{\boldmath$\Omega_{m}     $} & $0.271^{+0.022+0.048+0.072}_{-0.024-0.044-0.065}$\\

{\boldmath$\phi_{\Lambda, 1}$} & $-0.022^{+0.018+0.043+0.076}_{-0.023-0.040-0.055}$\\
\hline
\end{tabular}
    \caption{Mean values of the parameters $H_0$ (km s$^{-1}$Mpc$^{-1}$), $\Omega_m$ and $\phi_{\Lambda,1}$ and 68.3\%, 95.4\% and 99.7\% c.l. limits for Case III, $\sigma=3$, with SH0ES prior over $H_0$.}
    \label{tab04}
\end{table}


Comparison of the values in Tables \ref{tab03} and \ref{tab04} shows that the $H_0$ value from the analysis with SH0ES prior is larger than the $H_0$ value coming from the analysis with Planck prior. Conversely, the value of the parameter $\phi_{\Lambda,1}$ from SH0ES is smaller than the corresponding value from Planck. In fact, the result $\phi_{\Lambda,1}=-0.022^{+0.018+0.043}_{-0.023-0.040}$ from the analysis with SH0ES prior indicates that the constancy of $\Lambda$ is only attained at more than 1$\sigma$ c.l., although it is achieved below 2$\sigma$ c.l..We may say, then, that this result is only marginally compatible with the constancy of $\Lambda$.

\begin{figure}[h!]
    \centering
    \includegraphics[width=.5\linewidth]{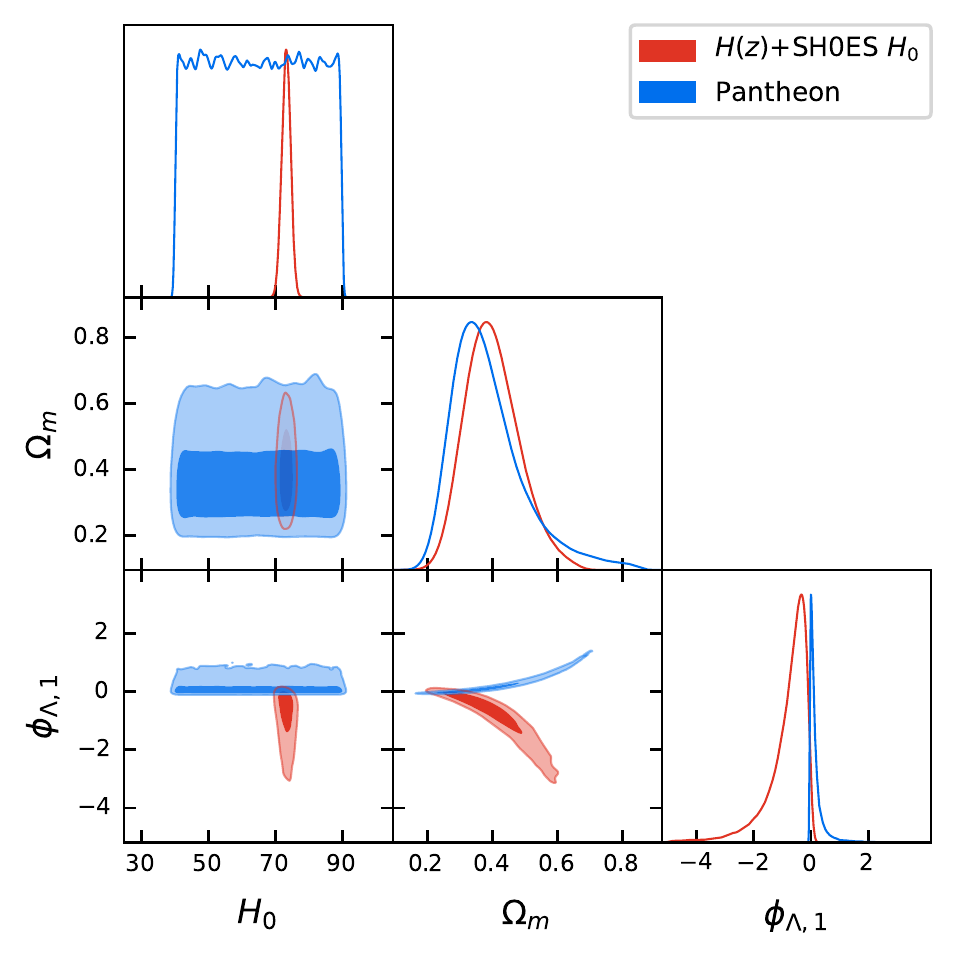}
    \caption{Constraints on the parameters $\{H_0,\Omega_{m,0},\phi_{\Lambda,1}\}$ in \textbf{Case III (B)} of the co-varying-$\{c,G,\Lambda\}$ model from two sets of data: (i) $H(z)$ data + $H_0$ prior from SH0ES data (red); and (ii) Pantheon data (blue). The statistical contours show 68.3\% and 95.4\% c.l. regions. }
    \label{HzH0Riess}
\end{figure}

\begin{figure}[h!]
    \centering
    \includegraphics[width=.5\linewidth]{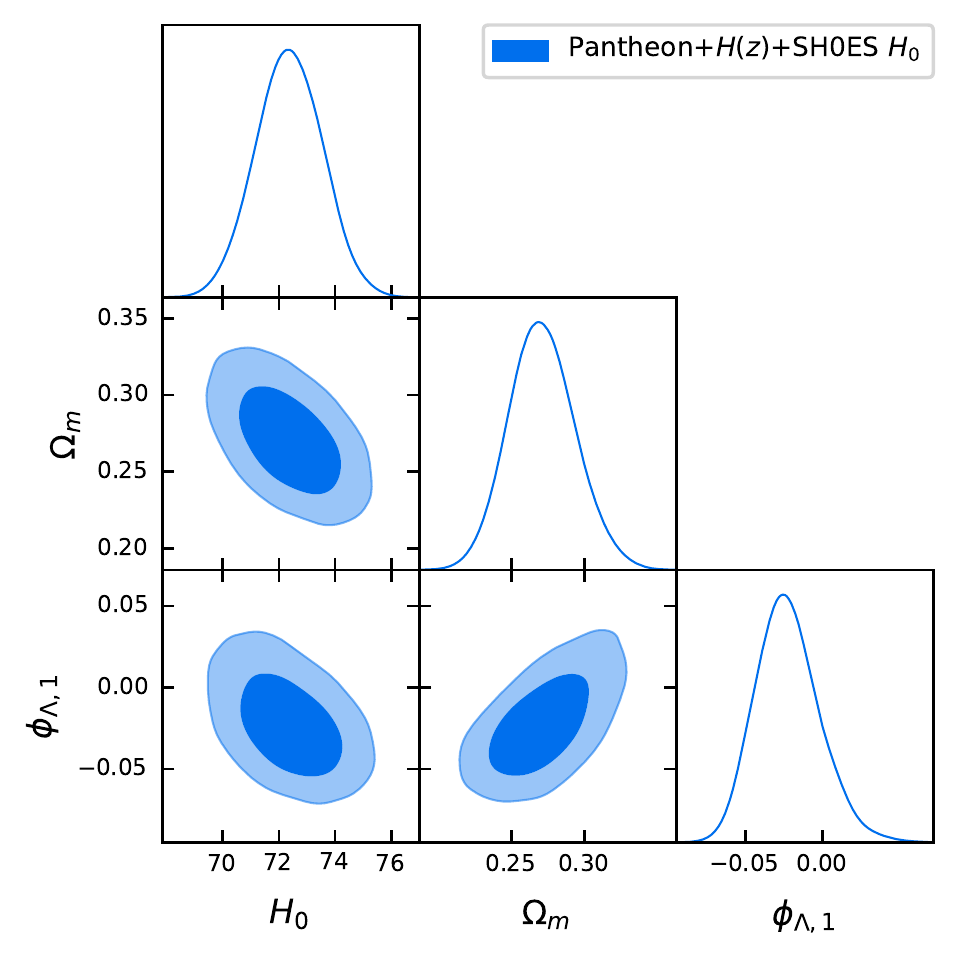}
    \caption{Constraints on the parameters $\{H_0,\Omega_{m,0},\phi_{\Lambda,1}\}$ in \textbf{Case III (B)} of the co-varying-$\{c,G,\Lambda\}$ model from $H(z)$ data + SH0ES $H_0$ + Pantheon.  The statistical contours show 68.3\% and 95.4\% c.l. regions. }
    \label{HzH0Riesscomb}
\end{figure}

Fig.~\ref{HzH0Riess} exhibits the statistical contours for the constraining analysis of our Co-varying-$\Lambda$ Dark Energy in the face of $H(z)$ data and SNe Ia data, both executed with the adoption of SH0ES $H_0$ prior. In this figures, the two data sets are shown separately. Figure \ref{HzH0Riesscomb} displays the combination of these two data sets. Now we are prepared to look at Case III (A) and Case III (B) in perspective. The corresponding figures are Fig.~\ref{HzPlanckH0comb} and Fig.~\ref{HzH0Riesscomb}. From the plots in these figures we infer that there is an anti-correlation between $H_0$ and $\phi_{\Lambda,1}$. This explains the fact that the larger value of $H_0$ estimated from SH0ES leads to a smaller value for $\phi_{\Lambda,1}$.


\section{Discussion} 
\label{sec:Discussion}


In this paper we have considered a cosmological model admitting the simultaneous variation of the fundamental couplings $G$, $c$, and $\Lambda$. The model was constrained using $H(z)$ data and SNe Ia data. The predicted values for the $H_0$ and $\Omega_{m,0}$ within this model are consistent with those derived from the $\Lambda$CDM model. Our model bears two additional parameters, $\phi_{\Lambda,1}$ and $\sigma$, both related to the varying character of the couplings $\{G,c,\Lambda\}$. Parameter $\sigma$ controls the way that $G$ scales in terms of $c$; in fact, $(G/G_0) = (c/c_0)^{\sigma}$---see Eq.~(\ref{eq:GdotOverG-GuptaAnsatz}). Parameter $\phi_{\Lambda,1}$ controls the amount by which cosmological term $\Lambda$ deviates from the constant value $\Lambda_0$. The data constraining process showed that our co-varying-$\{G,c,\Lambda\}$ model is able to fit the data successfully; this fact, justifies to call it Co-varying-$\Lambda$ Dark Energy since the present day accelerated dynamics is accommodated within the model. However, the data fitting also pointed to a $\phi_{\Lambda,1}$ value consistent with zero. This would favor non-varying $\Lambda$. Since the $z$-dependent part of the varying speed of light is multiplied by $\phi_{\Lambda,1}$---cf. Eqs.~(\ref{eq:c(phi_c)_Lambda(phi_Lambda)}) and (\ref{eq:CPC-phi_c}), this means that the data favor $c=c_0=$ constant, ultimately leading to $G=G_0=$ constant. In the latter sense, our Co-varying-$\Lambda$ Dark Energy recovers the ordinary dark energy description of $\Lambda$CDM model (with constant couplings). Moreover, parameter $\sigma$ remains unconstrained after our data fitting; it could not be confirmed to assume the value $\sigma=3$, as expected from other papers (e.g. \cite{cuzinatto2022dynamical,eaves2021constraints,gupta2020cosmology,Gupta:2020wgz,gupta2022faint,gupta2022effect,gupta2022constraining}).

After distilling the essentials of the results, let us put them in perspective, discuss them further, and indicate the importance of our model both from the theoretical standpoint and from the data-constraining point of view. 

When one contemplates the possibility of varying fundamental constants, such as $G$, $c$ and $\Lambda$, one is faced with the task of specifying the functional form for the spacetime variations of these couplings. This task seems to allow for an arbitrarily large number of possible independent function for $G(x^{\mu})$, $c(x^{\mu})$, and $\Lambda(x^{\mu})$. This arbitrariness is  undesirable due to its lacking of a guiding  underlying fundamental principle. Our Co-varying Physical Couplings (CPC) framework offers a scheme for entangling the variations of the couplings $\{G,c,\Lambda \}$ from very reasonable assumptions on the way one describes the gravitational interaction. In fact, these assumptions are the same as those in general relativity (GR): the spacetime four-dimensional manifold is metric compatible, the stress-energy tensor is covariantly conserved, Bianchi identity holds. These assumptions lead to the General Constraint (GC) which (i) forces the couplings $\{G,c,\Lambda \}$ to vary together, and (ii) reduces the arbitrariness of possible choices for the functions  $G(x^{\mu})$, $c(x^{\mu})$, and $\Lambda(x^{\mu})$. The CPC framework is also a minimal generalization of GR in the sense that the Einstein field equations (EFE) for $g_{\mu \nu}$ are kept intact. Our paper showed that the EFE and the GC lead to equations for background cosmology from which one can deduce the functional form of $c(z)$, once an ansatz for $\Lambda(z)$ is provided. This is a key advantage of our set up, because a reasonable proposal for the function $\Lambda(z)$ can be justified from late-time universe observational probes, such as the ones used here. Gupta's relation, $\dot{G}/G = \sigma \left( \dot{c}/c \right)$, then completes the scheme by enabling to obtain $G(z)$ from the equation for $c(z)$. To summarize, by proposing
\begin{equation}
\frac{\Lambda}{\Lambda_0} = 1+\phi_{\Lambda,1}\frac{z}{\left(1+z\right)}
\label{eq:CPC-Lambda(z)}
\end{equation}
we evaluate
\begin{equation}
\begin{cases}
\frac{c}{c_{0}}=\left\{ 1-\frac{\left(1-\Omega_{m,0}\right)}{\Omega_{m,0}}\frac{\phi_{\Lambda,1}}{4}\left[1-\left(1+z\right)^{-4}\right]\right\} ^{-\frac{1}{\left(4-\sigma\right)}}\\
\\
\frac{G}{G_{0}}=\left(\frac{c}{c_{0}}\right)^{\sigma}
\end{cases}.\label{eq:CPC-c(z)-G(z)}
\end{equation}
The equations above constitute the Co-varying-$\Lambda$ model. It involves all the couplings appearing naturally in the field equations of the gravitational interaction. In a broader picture, one could consider other couplings to be varying, such as the fine structure constant $\alpha$~(e.g. \cite{colacco2019galaxy,gonccalves2020variation,leefer2013new}), the Planck constant $\hbar$ and the Boltzmann constant $k_B$
~(e.g. \cite{gupta2020cosmology,gupta2021constraint,gupta2021testing,Gupta:2020wgz,gupta2022faint,gupta2022varying,gupta2022effect,gupta2022constraining,lee2021minimally}). Herein, we do not claim that our results should give the same results as in these references. Different initial hypothesis lead to different results, in general. Therefore it might as well be that models involving varying $\{ G, c, \Lambda, \hbar, k_B\}$ favor varying fundamental couplings after data fitting. In our case, after fitting the Co-varying-$\Lambda$ model to data, our conclusion is that the standard picture of $\{G,c,\Lambda\}$ as genuine constants is favored. We turn to the discussion of the data fitting process now.

The observational data used to constrain the possible simultaneous variations of $G$, $c$, and $\Lambda$ within our model were the 31 differential age $H(z)$ data \cite{Magana:2017nfs} and the 1048 Supernovae type Ia data from the Pantheon sample \cite{Pan-STARRS1:2017jku}. The use of both of these methods depend on the expression for the luminosity distance $d_L$. The equation for $d_L$ was modified to account for extra terms coming from the co-varying couplings $\{G,c,\Lambda\}$. This task was carefully undertaken in Section \ref{subsec:DistancesFluxLuminosity}, where the concepts of flux, magnitudes, and distance modulus were discussed in the light of the CPC framework.

In Case I (Section \ref{subsec:minimal-CPC-model}), we have particularized our analysis to the minimal CPC framework, wherein only $\left\{ c,G\right\}$ are supposed to vary, leaving the cosmological constant $\Lambda$ fixed. As previously mentioned, the gravitational coupling $G$ and the speed of light $c$ are related by $G \propto c^4$ in this context. Moreover we considered the parameterizarions P I and P II in Eq.~(\ref{eq:c(z)}) for $c$ as a function of the redshift $z$. Both P I and P II include $c_1$, the dimensionless parameter controlling the deviation of $c(z)$ from the constant value $c_0 = 299\,792\,458$ m/s. The data fitting values found for the deviation parameter $c_1$ are in full agreement with no variation of the speed of light w.r.t the redshift. The results with 95\% c.l. were $c_1 = 0.005^{+0.043}_{-0.042}$ for P I and $c_1 = 0.022^{+0.098}_{-0.091}$ for P II. These values  are in agreement to the ones found in our previous work \cite{cuzinatto2022observational}. Therein, we used galaxy cluster gas mass fraction data set to constrain $c_1 = -0.025 \pm 0.027$ and $c_1 = -0.037 \pm 0.038$ for P I and P II, respectively. Herein, we were also able to estimate values for the cosmological parameters $H_0$ and $\Omega_m$, which are found to be in good agreement with the last Planck 2018 \cite{Planck:2018vyg} results (see Table \ref{tab01}).  

The full CPC model admitting simultaneous entangled variation of $\left\{ c,G,\Lambda\right\}$ according to Eqs.~(\ref{eq:CPC-Lambda(z)}) and (\ref{eq:CPC-c(z)-G(z)}) was considered in Section \ref{subsec:LinearLambda-CPC-Model}. Two separate sets of analyses were performed in this context. Case II (Section \ref{subsubsec:LinearLambda-NoPrior}) does not entail any prior over the value of $H_0$. Case III (Section \ref{subsubsec:LinearLambda-WithPrior}) assumes priors over $H_0$. 

Let us begin by commenting on Case II. In our first attempt within this case, our data fitting analysis was performed by keeping parameter $\sigma$ of the relation $G \propto c^{\sigma}$ free (with an unconstrained interval of possible values). The result found was that $\sigma$ can not be constrained by the $H(z)$ and SNe Ia observational data sets. In our second analysis within Case II we set $\sigma = 3$ based on the works by Gupta and others (e.g.  \cite{cuzinatto2022dynamical,eaves2021constraints,gupta2020cosmology,Gupta:2020wgz,gupta2022faint,gupta2022effect,gupta2022constraining}) suggesting that this value is the preferred one, both from the theoretical aspect and from the phenomenological side. For this particular choice of $\sigma$, the parameter $\phi_{\Lambda,1}$ quantifying the deviation from a fixed cosmological constant was constrained to $\phi_{\Lambda,1} = 0.039^{+0.12}_{-0.095}$ at 95\% c.l.. This points out to no variation of $\Lambda$ in terms of $z$. The values estimated for $H_0$ and $\Omega_m$ in the context of Case II are also in good agreement to Planck 2018 results (see Table \ref{tab02}).

In Case III of the full CPC model, we have tested two priors over $H_0$, namely, the local $H_0$ from SH0ES and the early-universe $H_0$ from Planck. As there was a negative correlation between $H_0$ and $\phi_{\Lambda,1}$, and the Planck $H_0$ is lower than the SH0ES $H_0$, $\phi_{\Lambda,1}$ was slightly higher for Planck prior. However, the $\phi_{\Lambda,1}$ uncertainty was smaller with the SH0ES prior; it is precisely  in this case that the $\Lambda$ constancy is more challenged, being reached only at more than 1 $\sigma$ c.l.

Both analysis in Case II and Case III indicate no variation of the fundamental constants $\left\{ c,G,\Lambda\right\}$ within the Co-varying-$\Lambda$ Dark Energy of CPC framework. On top of that, the $H(z)$ and SNe Ia data fitting to the minimal CPC model in Case I also indicate that $\left\{ c,G\right\}$ are genuine constants. However, it is important to emphasize that we have utilized observational windows of the relatively late-time universe. Therefore, our constraining to the CPC models are trustworthy in the corresponding period of the cosmological evolution. By this comment we mean that our data constraining favor constant $\left\{ c,G,\Lambda\right\}$ in the late time universe. We can not affirm that this would be the case during the early universe evolution. Early-universe observational data sets (such as CMB data) could be employed to try and assess the viability of the CPC framework in the past history of the universe.

Another important remark is the following. The data used in our constraining analysis was not re-calibrated to take into account any possible effects of the varying couplings $\left\{ c,G,\Lambda\right\}$ built in the process of getting observational data tables in \cite{Magana:2017nfs} and \cite{Pan-STARRS1:2017jku} from the raw data outputs by the instruments. For example, the SNe Ia luminosity depends of the Chandrasekhar mass of the exploding star, which is a function of $c$ and $G$ \cite{gupta2022effect}; this was not considered in this study. Even though a more complete and rigorous analysis would demand adding further corrections and possibly re-calibrating raw data, our results are reasonable. In fact, the full agreement of the values we found for the parameters $H_0$ and $\Omega_m$ with the corresponding standard model ones hints that the CPC framework does not drastically influence the process of turning the raw observational data into the data values for $\mu$, etc. readily available in the literature. 

Other parametrizations and models of co-varying $\left\{ c,G,\Lambda\right\}$ are current under investigation. We are also taking into account different sets of observational data in order to further test the CPC framework as a possible viable generalization of GR.


\section*{Acknowledments}

RRC is grateful to Prof. Rajendra P. Gupta (University of Ottawa) for the hospitality and to CNPq-Brazil (Grant 309984/2020-3) for partial financial support. SHP acknowledges financial support from  {Conselho Nacional de Desenvolvimento Cient\'ifico e Tecnol\'ogico} (CNPq)  (No. 303583/2018-5 and 308469/2021-6).

\section*{Data Availability}


All data used in this paper are from the references cited.

\section*{Conflict of Interest}
Authors have no conflict of interest.



\begin{thebibliography}{}






\bibitem{abdalla2022cosmology}
Abdalla E.,  et~al., 2022, JHEAp, 34, 49

\bibitem{FermiGBMLAT:2009nfe}
Ackermann M.,  et~al., 2009, Nature, 462, 331

\bibitem{Planck:2018vyg}
Aghanim N.,  et~al., 2020, A\&A, 641, A6

\bibitem{Agrawal:2021cim}
Agrawal R.,  Singirikonda H.,   Desai S.,  2021, J. Cosmol. Astropart. Phys., 2021, 029

\bibitem{aich2022phenomenological}
Aich A.,  2022, Class. Quant. Grav., 39, 035010

\bibitem{Albrecht:1998ir}
Albrecht A.,  Magueijo J.,  1999, Phys. Rev. D, 59, 043516

\bibitem{Allen:2007ue}
Allen S.~W.,  Rapetti D.~A.,  Schmidt R.~W.,  Ebeling H.,  Morris G.,   Fabian
  A.~C.,  2008, MNRAS, 383, 879

\bibitem{Allen:2011zs}
Allen S.~W.,  Evrard A.~E.,   Mantz A.~B.,  2011, Ann. Rev. Astron.
  Astrophys., 49, 409

\bibitem{amendola2010dark}
Amendola L.,  Tsujikawa S.,  2010, Dark Energy: Theory and Observations.
Cambridge Univ. Press, Cambridge

\bibitem{Amendola:2006dg}
Amendola L.,  Camargo~Campos G.,   Rosenfeld R.,  2007, Phys. Rev. D, 75, 083506

\bibitem{anagnostopoulos2022swiss}
Anagnostopoulos F.~K.,  Bonanno A.,  Mitra A.,   Zarikas V.,  2022, Phys. Rev. D, 105, 083532




\bibitem{bailin1987kaluza}
Bailin D.,  Love A.,  1987, Rept. Prog. Phys., 50, 1087

\bibitem{ballardini2021cosmological}
Ballardini M.,  Finelli F.,   Sapone D.,  2022, J. Cosmol. Astropart. Phys., 06, 004

\bibitem{bamba2012dark}
Bamba K.,  Capozziello S.,  Nojiri S.,   Odintsov S.~D.,  2012, Astrophys. Space Sci., 342, 155

\bibitem{Barrow:1999is}
Barrow J.~D.,  1999, Phys. Rev. D, 59,  043515

\bibitem{Battaglia:2011cn}
Battaglia N.,  Bond J.~R.,  Pfrommer C.,   Sievers J.~L.,  2012, ApJ, 758, 74

\bibitem{baumann2009tasi}
Baumann D.,  2011, in {Theoretical Advanced Study Institute in Elementary Particle Physics}: {Physics of the Large and the Small}. p. 523

\bibitem{SDSS:2014iwm}
Betoule M.,  et~al., 2014, A\&A, 568, A22

\bibitem{bojowald2008loop}
Bojowald M.,  2005, Living Rev. Rel., 8, 11

\bibitem{bonanno2021effective}
Bonanno A.,  Kofinas G.,   Zarikas V.,  2021, Phys. Rev. D, 103, 104025

\bibitem{bora2021constraints}
Bora K.,  Desai S.,  2021, J. Cosmol. Astropart. Phys., 02,  012

\bibitem{brans1961mach}
Brans C.,  Dicke R.~H.,  1961, Phys. Rev., 124, 925

\bibitem{bull2016beyond}
Bull P.,  et~al., 2016, Phys. Dark Univ., 12, 56




\bibitem{Caldwell:1997ii}
Caldwell R.~R.,  Dave R.,   Steinhardt P.~J.,  1998, Phys. Rev. Lett., 80, 1582

\bibitem{callan1985strings}
Callan Jr. C.~G.,  Martinec E.~J.,  Perry M.~J.,   Friedan D.,  1985, Nucl. Phys. B, 262, 593

\bibitem{Cao:2016dgw}
Cao S.,  Biesiada M.,  Jackson J.,  Zheng X.,  Zhao Y.,   Zhu Z.-H.,  2017, J. Cosmol. Astropart. Phys., 2017, 012

\bibitem{Cao:2018rzc}
Cao S.,  Qi J.,  Biesiada M.,  Zheng X.,  Xu T.,   Zhu Z.-H.,  2018, ApJ, 867, 50

\bibitem{carlip2001quantum}
Carlip S.,  2001, Rept. Prog. Phys., 64,  885

\bibitem{Carroll:2004st}
Carroll S.~M.,  2019, {Spacetime and Geometry}.
Cambridge Univ. Press, Cambridge

\bibitem{chakrabarti2022generalized}
Chakrabarti S.,  2022, MNRAS, 513, 1088

\bibitem{colacco2019galaxy}
Cola\c{c}o L.~R.,  Holanda R. F.~L.,  Silva R.,   Alcaniz J.~S.,  2019, J. Cosmol. Astropart. Phys., 2019, 014

\bibitem{Colgain:2021pmf}
Colg\'ain E.~O.,  Sheikh-Jabbari M.~M.,   Yin L.,  2021, Phys. Rev. D, 104, 023510

\bibitem{costa2019covariant}
Costa R.,  Cuzinatto R.~R.,  Ferreira E. M.~G.,   Franzmann G.,  2019, Int. J. Mod. Phys. D, 28, 1950119

\bibitem{Cruz:2012bwp}
Cruz C.~N.,  Faria A. C. A.~d.,  2012, Phys. Rev. D, 86, 027703

\bibitem{cuzinatto2016scalar}
Cuzinatto R.~R.,  de Melo C. A.~M.,  Medeiros L.~G.,   Pompeia P.~J.,  2016, Phys. Rev. D, 93, 124034

\bibitem{Cuzinatto:2021ttc}
Cuzinatto R.~R.,  De~Morais E.~M.,   Pimentel B.~M.,  2021, Phys. Rev. D, 103, 124002

\bibitem{cuzinatto2022dynamical}
Cuzinatto R.~R.,  Gupta R.~P.,   Pompeia P.~J.,  2022a, preprint (arXiv:2204.00119)

\bibitem{cuzinatto2022observational}
Cuzinatto R.~R.,  Holanda R. F.~L.,   Pereira S.~H.,  2022b, preprint (arXiv:2202.01371)




\bibitem{da2022fundamental}
Da~Fonseca V.,  et~al., 2022, preprint (arXiv:2204.02930)

\bibitem{DeSabbata:1986sv}
De~Sabbata V.,  Gasperini M.,  1986, {Introduction to Gravitation}.
World Scientific Press, Singapore

\bibitem{dirac1937cosmological}
Dirac P.~A.,  1937, Nature, 139,  323

\bibitem{di2021realm}
Di~Valentino E.,  et~al., 2021, Class. Quant. Grav., 38, 153001


\bibitem{Duff:2014mva}
Duff M.~J.,  2015, Contemp. Phys., 56,  35




\bibitem{eaves2021constraints}
Eaves R.,  2021, MNRAS, 505, 3590

\bibitem{ellis2005c}
Ellis G. F.~R.,  Uzan J.-P.,  2005, Am. J. Phys.,  73, 240

\bibitem{ellis2012relativistic}
Ellis G.~F.,  Maartens R.,   MacCallum M.~A.,  2012, Relativistic cosmology.
Cambridge Univ. Press, Cambridge

\bibitem{Ettori:2009wp}
Ettori S.,  Morandi A.,  Tozzi P.,  Balestra I.,  Borgani S.,  Rosati P.,
  Lovisari L.,   Terenziani F.,  2009, A\&A, 501, 61




\bibitem{faraoni2004cosmology}
Faraoni V.,  2004, Cosmology in Scalar-Tensor Gravity, 
Vol. 139. Kluwer, Dordrecht

\bibitem{Foreman-Mackey:2012any}
Foreman-Mackey D.,  Hogg D.~W.,  Lang D.,   Goodman J.,  2013, Publ. Astron. Soc. Pac., 125, 306

\bibitem{fradkin1985quantum}
Fradkin E.~S.,  Tseytlin A.~A.,  1985, Nucl. Phys. B, 261, 1

\bibitem{franchino2021cosmological}
Franchino-Vi{\~n}as S.,  Mosquera M.,  2021, preprint (arXiv:2107.02243)

\bibitem{franzmann2017varying}
Franzmann G.,  2017, preprint (arXiv:1704.07368)




\bibitem{galli2013clusters}
Galli S.,  2013, Phys. Rev. D, 87,  123516

\bibitem{garcia2011upper}
Garcia-Berro E.,  Loren-Aguilar P.,  Torres S.,  Althaus L.~G.,   Isern J.,  2011, J. Cosmol. Astropart. Phys., 2011, 021

\bibitem{gonccalves2020variation}
Gon\c{c}alves R.~S.,  Landau S.,  Alcaniz J.~S.,   Holanda R. F.~L.,  2020, J. Cosmol. Astropart. Phys., 2020, 036

\bibitem{goodman2010communications}
Goodman J.,  Weare J.,  2010, Commun. Appl. Math. Comput. Sci., 5, 65

\bibitem{gupta2020cosmology}
Gupta R.~P.,  2020, MNRAS, 498, 4481

\bibitem{gupta2021constraint}
Gupta R.~P.,  2021a, Res. Notes AAS, 5, 30

\bibitem{gupta2021testing}
Gupta R.~P.,  2021b, Res. Notes AAS, 5,  176

\bibitem{Gupta:2020wgz}
Gupta R.~P.,  2021c, Astropart. Phys., 129, 102578

\bibitem{gupta2022faint}
Gupta R.~P.,  2021d, MNRAS, 509, 4285

\bibitem{gupta2022varying}
Gupta R.~P.,  2022a, preprint (arXiv:2201.11667)

\bibitem{gupta2022effect}
Gupta R.~P.,  2022b, MNRAS, 511, 4238

\bibitem{gupta2022constraining}
Gupta R.~P.,  2022c, MNRAS, 513, 5559




\bibitem{hanimeli2022can}
Han\i{}meli E.~T.,  Tutusaus I.,  Lamine B.,   Blanchard A.,  2022, Universe, 8, 148

\bibitem{heisenberg2022simultaneously}
Heisenberg L.,  Villarrubia-Rojo H.,   Zosso J.,  2022, preprint (arXiv:2201.11623)

\bibitem{Holanda:2020sqm}
Holanda R. F.~L.,  Pordeus-da Silva G.,   Pereira S.~H.,  2020, J. Cosmol. Astropart. Phys., 2020, 053




\bibitem{jofre2006constraining}
Jofre P.,  Reisenegger A.,   Fernandez R.,  2006, Phys. Rev. Lett., 97, 131102

\bibitem{joseph2022cosmology}
Joseph G.~W.,  \"Ovg\"un A.,  2022, Indian J. Phys., 96, 1861




\bibitem{king2012spatial}
King J.~A.,  Webb J.~K.,  Murphy M.~T.,  Flambaum V.~V.,  Carswell R.~F.,
  Bainbridge M.~B.,  Wilczynska M.~R.,   Koch F.~E.,  2012, MNRAS, 422, 3370

\bibitem{klypin1999missing}
Klypin A.~A.,  Kravtsov A.~V.,  Valenzuela O.,   Prada F.,  1999, ApJ, 522, 82

\bibitem{kotuvs2017high}
Kotu\v{s} S.~M.,  Murphy M.~T.,   Carswell R.~F.,  2017, MNRAS, 464, 3679




\bibitem{lazaridis2009generic}
Lazaridis K.,  et~al., 2009, MNRAS, 400, 805

\bibitem{lee2021determination}
Lee S.,  2021a, preprint (arXiv:2110.08809)

\bibitem{lee2021cosmic}
Lee S.,  2021b, preprint (arXiv:2108.06043)

\bibitem{lee2021minimally}
Lee S.,  2021c, J. Cosmol. Astropart. Phys., 2021, 054

\bibitem{leefer2013new}
Leefer N.,  Weber C. T.~M.,  Cing\"oz A.,  Torgerson J.~R.,   Budker D.,  2013, Phys. Rev. Lett., 111, 060801

\bibitem{lee2022local}
Lee B.-H.,  Lee W.,  Colg\'ain E.~O.,  Sheikh-Jabbari M.~M.,   Thakur S.,  2022, J. Cosmol. Astropart. Phys., 2022, 004

\bibitem{leon2022inflation}
Leon G.,  2022, Class. Quant. Grav., 39,  075008

\bibitem{liu2018light}
Liu Y.,  Ma B.-Q.,  2018, Eur. Phys. J. C, 78, 825

\bibitem{Liu:2021eit}
Liu T.,  Cao S.,  Biesiada M.,  Liu Y.,  Lian Y.,   Zhang Y.,  2021a, MNRAS, 506, 2181

\bibitem{liu2021probing}
Liu Z.-E.,  Liu W.-F.,  Zhang T.-J.,  Zhai Z.-X.,   Bora K.,  2021b, ApJ, 922, 19




\bibitem{maeda2022cuscuta}
Maeda K.-I.,  Panpanich S.,  2022, Phys. Rev. D, 105, 104022

\bibitem{Magana:2017nfs}
Magana J.,  Amante M.~H.,  Garcia-Aspeitia M.~A.,   Motta V.,  2018, MNRAS, 476, 1036

\bibitem{Mantz:2014xba}
Mantz A.~B.,  Allen S.~W.,  Morris R.~G.,  Rapetti D.~A.,  Applegate D.~E.,  Kelly P.~L.,  von~der Linden A.,   Schmidt R.~W.,  2014, MNRAS, 440, 2077

\bibitem{martins2021varying}
Martins C. J. A.~P.,  2021, 
preprint (arXiv:2111.04137)

\bibitem{martins2022varying}
Martins C. J. A.~P.,  Ferreira F. P. S.~A.,   Marto P.~V.,  2022, Phys. Lett. B, 827, 137002

\bibitem{mendoncca2021search}
Mendon\c{c}a I. E. C.~R.,  Bora K.,  Holanda R. F.~L.,  Desai S.,   Pereira
  S.~H.,  2021, J. Cosmol. Astropart. Phys., 2021, 034

\bibitem{Moffat:1992ud}
Moffat J.~W.,  1993, Int. J. Mod. Phys. D, 2, 351




\bibitem{ooba2016planck}
Ooba J.,  Ichiki K.,  Chiba T.,   Sugiyama N.,  2016, Phys. Rev. D, 93, 122002

\bibitem{overduin1997kaluza}
Overduin J.~M.,  Wesson P.~S.,  1997, Phys. Rept., 283, 303




\bibitem{Petit:1988ti}
Petit J.~P.,  1988, Mod. Phys. Lett. A,  3, 1527

\bibitem{Petit:1995ys}
Petit J.~P.,  1995, Astrophys. Space Sci., 226,  273

\bibitem{Petit:2008eb}
Petit J.~P.,  d'Agostini G.,  2008, preprint (arXiv:0803.1362)

\bibitem{pitjeva2021estimates}
Pitjeva E.~V.,  Pitjev N.~P.,  Pavlov D.~A.,   Turygin C.~C.,  2021, A\&A, 647, A141




\bibitem{Qi:2014zja}
Qi J.-Z.,  Zhang M.-J.,   Liu W.-B.,  2014, Phys. Rev. D, 90, 063526




\bibitem{Riess:2020fzl}
Riess A.~G.,  Casertano S.,  Yuan W.,  Bowers J.~B.,  Macri L.,  Zinn J.~C.,  Scolnic D.,  2021, ApJL,  908, L6

\bibitem{ryden2017introduction}
Ryden B.,  2017, Introduction to cosmology.
Cambridge Univ. Press, Cambridge




\bibitem{sakr2022can}
Sakr Z.,  Sapone D.,  2022, J. Cosmol. Astropart. Phys., 2022,  034

\bibitem{Salzano:2016hce}
Salzano V.,  2017, Phys. Rev. D, 95,  084035
  

\bibitem{Salzano:2014lra}
Salzano V.,  Dabrowski M.~P.,   Lazkoz R.,  2015, Phys. Rev. Lett., 114, 101304

\bibitem{Sarazin:1986zz}
Sarazin C.~L.,  1986, Rev. Mod. Phys., 58,  1

\bibitem{Pan-STARRS1:2017jku}
Scolnic D.~M.,  et~al., 2018, ApJ, 859, 101

\bibitem{Shojaie:2004xw}
Shojaie H.,  Farhoudi M.,  2006, Can. J. Phys., 84,  933

\bibitem{Shojaie:2004sq}
Shojaie H.,  Farhoudi M.,  2007, Can. J. Phys., 85,  1395

\bibitem{sonia2022dynamical}
Sonia C.,  Singh S.~S.,  2022, preprint (arXiv:2203.16892)

\bibitem{sotiriou2010f}
Sotiriou T.~P.,  Faraoni V.,  2010, Rev. Mod. Phys., 82, 451




\bibitem{uzan2003fundamental}
Uzan J.-P.,  2003, Rev. Mod. Phys., 75,  403

\bibitem{uzan2011varying}
Uzan J.-P.,  2011, Living Rev. Rel., 14, 2




\bibitem{van2015variation}
Van~de Bruck C.,  Mifsud J.,   Nunes N.~J.,  2015, J. Cosmol. Astropart. Phys., 2015, 018

\bibitem{verbiest2008precision}
Verbiest J. P.~W.,  et~al., 2008, ApJ,  679, 675

\bibitem{vijaykumar2021constraints}
Vijaykumar A.,  Kapadia S.~J.,   Ajith P.,  2021, Phys. Rev. Lett., 126, 141104




\bibitem{Weinberg:1972kfs}
Weinberg S.,  1972, {Gravitation and Cosmology}: {Principles and Applications of the General Theory of Relativity}.
John Wiley and Sons, New York




\bibitem{Xu:2016zxi}
Xu H.,  Ma B.-Q.,  2016a, Astropart. Phys., 82, 72

\bibitem{xu2016light}
Xu H.,  Ma B.-Q.,  2016b, Phys. Lett. B, 760, 602




\bibitem{zhao2018constraining}
Zhao W.,  Wright B.~S.,   Li B.,  2018, J. Cosmol. Astropart. Phys. 2018, 052

\bibitem{zhu2021pre}
Zhu J.,  Ma B.-Q.,  2021, Phys. Lett. B, 820, 136518


\end{thebibliography}
\end{document}